\documentclass[superscriptaddress,groupedaddress,nofootnoteinbib,11pt]{article}
\usepackage{jheppub}

\usepackage[utf8]{inputenc}
\usepackage{amsfonts,amsmath,amssymb,amsthm,graphicx,hyperref,mathrsfs,xcolor}
\usepackage{caption}

\allowdisplaybreaks[3]

\def\({\left(}
\def\){\right)}
\def\[{\left[}
\def\]{\right]}
\def\d{\mathrm{d}}
\newcommand{\n} {\nabla}
\newcommand{\p} {\partial}
\newcommand{\f}[2]{\frac{#1}{#2}}
\def \bal#1\eal  {\begin{align} #1 \end{align}}
\newcommand{\eref}[1]{{Eq.~(\ref{#1})}}
\newcommand{\be} {\begin{equation}}
\newcommand{\ee} {\end{equation}}
\newcommand{\bc}{\begin{center}}
\newcommand{\ec}{\end{center}}
\newcommand{\bim} {\begin{itemize}[noitemsep]}

\newcommand{\eim} {\end{itemize}}

\newcommand{\mc} {\mathcal}

   \newcommand{\bfk} {{\bf k}}   \newcommand{\bfl} {{\bf l}}

      \newcommand{\bfx} {{\bf x}}

\newcommand{\mn} {{\mu\nu}}

\newcommand{\ai}{\alpha} 
\newcommand{\di}{\delta} 
\newcommand{\thi}{\theta}   
\newcommand{\li}{\lambda} 
\newcommand{\cph}{\varphi} 
\newcommand{\oi}{\omega} 
\newcommand{\Di}{\Delta} 

\newcommand{\lan}{\langle}
\newcommand{\ran}{\rangle}
\renewcommand{\t}{\tilde}

\graphicspath{ {draft-images/} }

\title{Quantum corrected Q-ball dynamics}

\author[a]{Qi-Xin Xie}
\author[b]{, Paul M. Saffin}
\author[c]{, Anders Tranberg}
\author[a,d,e]{, Shuang-Yong Zhou}

\affiliation[a]{Interdisciplinary Center for Theoretical Study, University of Science and Technology of China,
Hefei, Anhui 230026, China}
\affiliation[b]{School of Physics and Astronomy, University of Nottingham, University Park, Nottingham NG7 2RD, United Kingdom}
\affiliation[c]{Faculty of Science and Technology, University of Stavanger, 4036 Stavanger, Norway}
\affiliation[d]{Peng Huanwu Center for Fundamental Theory, Hefei, Anhui 230026, China}
\affiliation[e]{Theoretical Physics, Blackett Laboratory, Imperial College, London, SW7 2AZ, UK}

\emailAdd{xqx2018@mail.ustc.edu.cn}
\emailAdd{paul.saffin@nottingham.ac.uk}
\emailAdd{anders.tranberg@uis.no}
\emailAdd{zhoushy@ustc.edu.cn}

\preprint{{\small USTC-ICTS/PCFT-23-37}}

\abstract{The physics of individual Q-balls and interactions between multiple Q-balls are well-studied in classical numerical simulations. Interesting properties and phenomena have been discovered, involving stability, forces, collisions and swapping of charge between different components of multi-Q-ball systems. We investigate these phenomena in quantum field theory, including quantum corrections to leading order in a 2PI coupling expansion, the inhomogeneous Hartree approximation. 
The presence of quantum modes and new decay channels allows the mean-field Q-ball to exchange charge with the quantum modes, and also alters the charge swapping frequencies of the composite Q-balls. It is also observed that the periodic exchanges between the mean-field and quantum modes tend to be quenched by collisions between Q-balls. We illustrate how the classical limit arises through a scaling of the Q-ball potential, making quantum corrections negligible for large-amplitude Q-balls.}

\begin{document}
\maketitle
\flushbottom

\section{Introduction}
\label{sec:introduction}

Many non-linear field theories give rise to solitonic solutions, characterized by spatially localized configurations with intriguing non-perturbative properties. Solitons may be stable for a variety of reasons. Whereas topological solitons derive their stability from topological charges and are typically static in time, Q-balls are non-topological solitons stabilized by the presence of conserved Noether charges, and are time-dependent, although typically periodic \cite{Friedberg:1976me,Coleman:1985ki}. Oscillons are quasi-periodic, non-topological and have no charge, but oscillate with a frequency lower than the particle mass, and so decay very slowly \cite{Bogolyubsky:1976yu, Copeland:1995fq}. Q-balls may arise under quite general conditions including the case of self-interacting complex scalar field theories, where the potential grows more slowly than the quadratic mass term away from its minimum so as to exhibit an attractive force that condenses charges into Q-balls.  The properties of Q-balls have been extensively investigated \cite{Battye:2000qj,Axenides:1999hs,Tsumagari:2008bv,Copeland:2014qra,Tamaki:2014oha,Kovtun:2018jae,Lennon:2021uqu,Kasuya:2022cko,Ansari:2023cay,Almumin:2023wwi}, and also in supersymmetric models \cite{Kusenko:1997zq,Enqvist:1997si}. They may arise naturally in the early stages of the universe \cite{Multamaki:2002hv,Tsumagari:2009na,Griest:1989cb,Hiramatsu:2010dx,Zhou:2015yfa,Lloyd-Stubbs:2021xlk} and are candidates for dark matter \cite{Kusenko:1997si,Kusenko:1997vp,Kusenko:2001vu,Krylov:2013qe,Ponton:2019hux,Bai:2019ogh}. Q-balls, along with their counterparts in strong gravitational environments known as boson stars \cite{Kaup:1968zz,Liebling:2012fv,Visinelli:2021uve} can induce superradiance phenomena \cite{Saffin:2022tub,Gao:2023gof,Cardoso:2023dtm}. A similar phenomenon, more closely linked to parametric resonance, has also been seen in the emission of fermions in an oscillon background \cite{Saffin:2016kof}. Experimentally, Q-balls can arise in cold atom systems \cite{Enqvist:2003zb,Bunkov:2007fe}, and their properties studied in a controlled environment. 

When more than one Q-ball is present a variety of interaction phenomena may occur, including attraction, repulsion, charge transfer, fission and fusion \cite{Axenides:1999hs,Battye:2000qj,Bowcock:2008dn,Brihaye:2009yr,Siemonsen:2023hko}. Different oscillation phases and overall signs of the charges give rise to complex and diverse outcomes. For example, two Q-balls with identical charge will attract each other if their scalar condensates are in phase and repel if they are out of phase, accompanied by charge transfer if the phases are not completely aligned or anti-aligned. The signs of their charge also affect the directions of interaction forces. Q-balls can combine to form a series of composite structures, including charge-swapping Q-balls (CSQs) \cite{Copeland:2014qra,Xie:2021glp,Hou:2022jcd}. In CSQs, positive and negative charge coexist, with the charges swapping location as the CSQ evolves. While CSQs can have very long life-times, they eventually decay into oscillons \cite{Bogolyubsky:1976yu, Copeland:1995fq, Zhou:2013tsa, Kou:2019bbc, VanDissel:2020umg, Zhang:2021xxa,Kou:2021bij, Wang:2022rhk, GalvezGhersi:2023omd, vanDissel:2023zva, Evslin:2023qbv}. Although most studies focus on classical Q-balls, quantum dynamical aspects have been considered as well \cite{Graham:2001hr,Tranberg:2013cka}.

In this paper, we shall investigate quantum corrections to the dynamics of Q-balls. While the dynamics of a single Q-ball was considered in \cite{Tranberg:2013cka}, we generalise this to a numerical study of interactions among multiple Q-balls. Due to the time-dependent and spatially inhomogeneous nature of Q-balls, imaginary time Monte Carlo simulations do not apply, and we need to perform real-time quantum simulations. For this, we implement the inhomogeneous Hartree approximation \cite{Salle:2000jb,Salle:2003ju,Salle:2002fu,Salle:2000hd,Bettencourt:2001xg,Tranberg:2013cka,Saffin:2014yka,Borsanyi:2007wm}, which describes the dynamics using the one- and connected two-point functions only, neglecting all higher-order connected $n$-point functions of the quantum fluctuations. In this method, the one-point correlator plays the role of a classical inhomogeneous background field (the mean-field Q-ball). Such an inhomogeneous mean field interacts with the quantum field modes, opening up new channels for the Q-balls to potentially decay through, and also allows for a certain amount of scattering and intermediate-time quantum thermalisation \cite{Salle:2000hd}. The inhomogeneous Hartree approximation allows us to capture the leading order quantum effects in a loop expansion, taking fully into account the time-dependence and spatial inhomogeneity of the system. 

In a Gaussian approximation, including a free field and the Hartree approximation, the field operator may be expanded on a set of mode functions to be solved numerically. There is a time- and space-dependent mode function for each momentum mode ${\bf k}$, so that on a discrete spatial lattice of $N^d$ sites ($d$ being the number of the spatial dimensions), the numerical effort increases as $(N^d)^2$ (as opposed to a classical field simulation, which scales as $N^d$ only). The ensemble method \cite{Borsanyi:2007wm,Borsanyi:2008eu} \footnote{See also \cite{Berges:2010zv,Saffin:2011kc,Saffin:2011kn,Hebenstreit:2013qxa} for a number of applications.} reduces this issue by introducing an ensemble of configurations representing the initial quantum state, and evolving these in position space rather than momentum space. In effect, it allows for trading a non-stochastic simulation of $N^d$ mode functions ($N^d\times N^d$) for a simulation of an ensemble of $\mc{E}$ field configurations ($N^d\times \mc{E}$). If the required statistical precision of the ensemble-averaged observables is achieved for $\mc{E}<N^d$, this amounts to a more efficient procedure. This will be the approach we take in this paper. 

The paper is organized as follows. In Section \ref{sec:modelAndSetup}, we introduce the model and review some basic properties of classical Q-balls. We then introduce the inhomogeneous Hartree approximation for the quantum evolution, the stochastic ensemble average method, the renormalization scheme, and some details of the numerical implementation. In Section \ref{sec:singleQball}, we investigate the quantum corrections to the dynamics of a single Q-ball in 2+1D (an early exploratory work \cite{Tranberg:2013cka} considered 3+1D). Two regimes with substantial differences in quantum behavior are identified. In Section \ref{sec:multipleQball}, we study the interactions between Q-balls in various scenarios, including well separated Q-balls, collisions of Q-balls and charge-swapping Q-balls. We identify cases where the quantum simulations closely resemble the corresponding classical ones, and show examples of how the quantum corrections affect the multi-Q-ball dynamics for generic cases. We summarize in Section \ref{sec:summary}.

\section{Model and setup}
\label{sec:modelAndSetup}

In the following, we introduce the complex scalar field model we will be exploring, review the basic properties of classical Q-balls, and present the inhomogeneous Hartree approximation for including quantum corrections. We then specify the numerical setup used in the paper. 

\subsection{Classical Q-balls}
\label{sec:qball}

The simplest action that supports Q-ball solutions involves a single complex scalar field, $\varphi$, with a $U(1)$-symmetric potential with certain properties. We write 
\be
\label{eq:action}
S=\int \d^D x \big[-(\p^{\mu}\cph)^*(\p_{\mu}\cph)-V(|\cph|)\big].
\ee
We choose the metric convention $g_{\mn}=\text{diag}(-++\dots)$ for a $D=d+1$ dimensional Minkowski space-time. The attractive nature can be achieved by a potential that ``opens up'' away from the quadratic minimum \cite{Coleman:1985ki}. For simplicity, we will focus on the sextic polynomial potential 
\be
\label{eq:potential-sextic}
V(|\cph|)=m^2|\cph|^2-\li|\cph|^4+\f43g|\cph|^6,
\ee 
which could for instance arise as an effective low-energy theory once heavy degrees of freedom have been integrated out \cite{Friedberg:1976me}. The factor $4/3$ is chosen for convenience at the level of the equations of motion, which we will simulate both classically and quantum mechanically. The global $U(1)$ symmetry of the action (\ref{eq:action}), $\cph\to\cph^{i\thi}$ with $\thi$ being a real constant, implies the conservation of the Noether charge for an isolated system
\be 
\label{eq:charge}
Q=\int \d^d x  j^0=\int \d^d x  i(\cph^*\p_t\cph-\cph\p_t\cph^*)=\int \d^d x  (\phi_2\p_t\phi_1-\phi_1\p_t\phi_2),
\ee 
where $j^0$ is the charge density and we have defined the real and imaginary components of $\cph$
\be
\cph=\f1{\sqrt2}(\phi_1+i\phi_2) ,
\ee
with $\phi_{1,2}$ real-valued. The conserved energy of the system is given by
\be 
\label{eq:energy}
E=\int \d^d x  \(|\p_t\cph|^2+|\underline\n\cph|^2+V(|\cph|)\),
\ee 
and the classical equations of motion, in terms of $\phi_1$ and $\phi_2$, are 
\begin{subequations}
\label{eq:classicalEoM}
\bal 
[-\p_x^2+m^2-\li(\phi_1^2+\phi_2^2)+g(\phi_1^2+\phi_2^2)^2]\phi_1(x)&=0, \\
[-\p_x^2+m^2-\li(\phi_1^2+\phi_2^2)+g(\phi_1^2+\phi_2^2)^2]\phi_2(x)&=0.
\eal 
\end{subequations}
A classical Q-ball is a localized, minimum energy configuration for a fixed charge $Q$, whose profile has a spherical form with constant internal rotation in field space
\be 
\label{eq:ansatz-Qball}
\cph(x)=\f{1}{\sqrt2}f(r)e^{-i\oi t},
\ee 
where $r$ is the radial coordinate from the center of the Q-ball. The profile function $f(r)$ approaches a constant at $r=0$ and decays quickly away from the center. That is, $f(r)$ satisfies the boundary conditions $f'(0)=f(\infty)=0$. For the classical Q-ball, the potential energy of the field is determined by $f(r)$ only $V(f)=m^2f^2/2-\li f^4/4+g f^6/6$, and the $U(1)$ charge is 
\be
Q=\oi\int \d^d x f^2(r) .
\ee
For a classical Q-ball solution to exist, the absolute value of the frequency $|\oi|$ needs to be within the range $\oi_-\leq|\oi|<\oi_+$, where $\oi_-=(m^2-3\li^2/(16g))^{1/2}$ and $\oi_+=m$ \cite{Coleman:1985ki}. When $\oi<0$, the configuration is dubbed an anti-Q-ball, whose charge is negative. One may show that in 2+1D, Q-balls within this whole frequency range are stable against small perturbations, $\d Q/\d \oi\le0$, and satisfy the condition $E/Q<m$. For 3+1D, only a subset of this range supports classically stable Q-balls \cite{Tsumagari:2008bv}.

\subsection{The inhomogeneous Hartree approximation}
\label{sec:Hartree}

For a physical system where all the relevant momentum modes contain high occupation numbers, it is frequently argued that ensemble averages of classical field theory solutions provide reasonable approximations to the quantum dynamics \cite{Berges:2004yj}. In this paper, we further assess this point by incorporating quantum corrections in the Hartree approximation \cite{Salle:2000hd}. Because the background mean field is inhomogeneous, also the quantum corrections are inhomogeneous, as we will describe below. 

In the Hartree approximation, we only include information about the one-point and two-point correlation functions, while all higher order connected correlators are zero. Promoting $\phi_1$ and $\phi_2$ to be operators, we can re-interpret \eqref{eq:classicalEoM} as the Heisenberg equations of motion for quantum operators. They may also be understood as the lowest order of the Schwinger-Dyson equations. The one-point functions, or mean fields, are defined as
\bal
\label{eq:1point}
\Phi_1(x)=\lan\phi_1(x)\ran, \quad \Phi_2(x)=\lan\phi_2(x)\ran,
\eal 
and the connected two-point functions are defined as 
\begin{subequations}
\label{eq:2point}
\bal
G_1(x,y) &= \lan (\phi_1(x)-\Phi_1(x)) (\phi_1(y)-\Phi_1(y) )\ran =\lan \phi_1(x)\phi_1(y)\ran-\lan\phi_1(x)\ran\lan\phi_1(y)\ran, \\
G_2(x,y) &= \lan (\phi_2(x)-\Phi_2(x)) (\phi_2(y)-\Phi_2(y) )\ran =\lan\phi_2(x)\phi_2(y)\ran-\lan\phi_2(x)\ran\lan\phi_2(y)\ran, \\
K(x,y) &=  \lan (\phi_1(x)-\Phi_1(x)) (\phi_2(y)-\Phi_2(y) )\ran=\lan\phi_1(x)\phi_2(y)\ran-\lan\phi_1(x)\ran\lan\phi_2(y)\ran, \\
\bar{K}(x,y) &=  \lan (\phi_2(x)-\Phi_2(x)) (\phi_1(y)-\Phi_1(y) )\ran=\lan\phi_2(x)\phi_1(y)\ran-\lan\phi_2(x)\ran\lan\phi_1(y)\ran .
\eal 
\end{subequations}
When the two spacetime points are identical $x=y$, we write them as 
\be 
\label{eq:2point-abbr}
G_{1,2}(x,x)\equiv G_{1,2}, \quad K(x,x)= \bar{K}(x,x)\equiv K=\bar{K} .
\ee 
These correlators at coincidence contain divergences, and these are regularized by the lattice, as will be discussed later. 

Since in the Hartree approximation, high order connected correlators are set to zero 
\be
\lan \phi_{i_1}(x_1)\phi_{i_2}(x_2)... \phi_{i_n}(x_n)\ran_C = 0, ~~~~~n\geq 3 ,
\ee
where the subscript $C$ means the fully connected part. This leaves the disconnected parts, which can all be expressed in terms of $\Phi_{1,2}$ and $G_{1,2},K,\bar K$. This approximation clearly misses some physics, and for homogeneous systems adds only a state-dependent effective mass. Still, in particular for inhomogeneous systems it has been used to capture the leading quantum effects of many nonlinear systems \cite{Borsanyi:2008ar,Boyanovsky:1993pf,Bergner:2003au,Salle:2003ju,Borsanyi:2007wm,Tranberg:2013cka,Bettencourt:2001xg}. 

The $c$-number equations of motion for the one-point functions can be obtained by taking the quantum expectation values of the operator equations of motion and making use of the Hartree approximation:
\begin{subequations}
\label{eq:1point-eom}
\bal 
[-\p_x^2+M_{11}^2(x)]\Phi_1(x)+M_{12}^2(x)\Phi_2(x)=0, \\
[-\p_x^2+M_{22}^2(x)]\Phi_2(x)+M_{21}^2(x)\Phi_1(x)=0,
\eal 
\end{subequations}
where  
\begin{subequations}
\label{eq:1point-massMatrix}
\bal 
    M_{11}^2&=m^2-\li(\Phi_1^2+\Phi_2^2)+g(\Phi_1^4+\Phi_2^4+2\Phi_1^2\Phi_2^2) -\li(3G_1+G_2) +12gK(K+\Phi_1\Phi_2) \notag  \\
    & ~~~~ +g(15G_1^2+3G_2^2+6G_1G_2+10G_1\Phi_1^2+6G_1\Phi_2^2+6G_2\Phi_2^2+2G_2\Phi_1^2), \\
M_{12}^2&=-2\li K+4gK(\Phi_2^2+3G_1+3G_2),
\eal 
\end{subequations}
and $M_{21}^2,M_{22}^2$ can be obtained from $M_{12}^2,M_{11}^2$ by swapping indices between 1 and 2. It is easy to see that if we only keep the one-point functions, Eqs.~(\ref{eq:1point-eom}) reduce to the classical field equations. Similarly, the equations of motion of the two-point functions can be obtained by multiplying the Heisenberg field equations by $\phi_i(y)$ and then taking their quantum expectation values 
\begin{subequations}
\label{eq:2point-eom}
\bal
[-\p_x^2+\bar{M}_{11}^2(x)]G_1(x,y)+\bar{M}_{12}^2(x)\bar{K}(x,y)=0, \\
[-\p_x^2+\bar{M}_{11}^2(x)]K(x,y)+\bar{M}_{12}^2(x)G_2(x,y)=0, \\
[-\p_x^2+\bar{M}_{22}^2(x)]G_2(x,y)+\bar{M}_{21}^2(x)K(x,y)=0, \\
[-\p_x^2+\bar{M}_{22}^2(x)]\bar{K}(x,y)+\bar{M}_{21}^2(x)G_1(x,y)=0,
\eal 
\end{subequations}
where again the Hartree approximation is used and
\begin{subequations}
\label{eq:2point-massMatrix}
\bal 
    \bar{M}_{11}^2 & =m^2-\li(3\Phi_1^2+\Phi_2^2)+g(5\Phi_1^4+\Phi_2^4+6\Phi_1^2\Phi_2^2) -\li(3G_1+G_2) +12gK(K+2\Phi_1\Phi_2) \notag \\
    &~~~~ +g(15G_1^2+3G_2^2+6G_1G_2+30G_1\Phi_1^2+6G_1\Phi_2^2+6G_2\Phi_2^2+6G_2\Phi_1^2), \\
\bar{M}_{12}^2&=-2\li\Phi_1\Phi_2+4g(\Phi_1^3\Phi_2+\Phi_1\Phi_2^3+3G_1\Phi_1\Phi_2+3G_2\Phi_1\Phi_2) \notag \\
& ~~~~ -2\li K+12gK(\Phi_1^2+\Phi_2^2+G_1+G_2),
\eal 
\end{subequations}
with $\bar{M}_{21}^2,\bar{M}_{22}^2$ obtained by swapping the 1 and 2 indices. As the two-point functions in \eqref{eq:2point-eom} contain both $x$ and $y$, it is numerically expensive to evolve these equations with a finite difference method. In a homogeneous system (with no mean field or when the field is not space-dependent), $\Phi_{1,2}(x) =\Phi_{1,2}(t)$ all the $M$ and $\bar{M}$ are also only time-dependent, and two-point correlators satisfy $G_{11}(x,y) = G_{1}(t,t',|{\bf x-y}|)$ (and similar for the other components $G_2$, $K$, ...). Homogeneous Hartree systems are straightforward to evolve numerically \cite{Aarts:2000wi}.

We proceed instead by incorporating the inhomogeneous Q-ball solution into the picture. We have defined the two-point functions $G_{1,2},K,\bar K$ to be deviations of $\phi_j(x)$ from the mean field $\Phi_j(x)$, and we shall initially identify $\Phi_j(x)$ to be the Q-ball solution along with the identification that $\phi_j(x)-\Phi_j(x)$ is a Gaussian field
\be 
\phi_j(x)-\Phi_j(x)=\cph_j(x) = \int\widetilde{\d k}\[a_{\bfk}^jf_{\bfk}^j(x)+a_{\bfk}^{j\dag}f_{\bfk}^{j*}(x)\], \quad j=1,2,
\ee 
where $\widetilde{\d k}=\d^d k/((2\pi)^d2\oi_k)$ with $\oi_k=\sqrt{\bfk^2+m_r^2}$ and $m_r$ the renormalized mass to be defined shortly, and the annihilation and creation operators $a_{\bfk}^j$ and $a_{\bfk}^{j\dag}$ satisfy the canonical commutation relations 
\be 
\label{eq:commutation}
[a_{\bfk}^{j_1},a_{\bfl}^{j_2\dag}]=(2\pi)^d2\oi_k\di^d(\bfk-\bfl)\di_{j_1j_2},~~~~[a_{\bfk}^{j_1},a_{\bfl}^{j_2}]=[a_{\bfk}^{j_1\dag},a_{\bfl}^{j_2\dag}]=0.
\ee 
It can then be shown that the mode functions $f_{\bfk}^j$ satisfy the following homogeneous equations of motion 
\begin{subequations}
\label{eq:mode-eom}
\bal
[-\p_x^2+\bar{M}_{11}^2(x)]f_{\bfk}^1(x)+\bar{M}_{12}^2(x)f_{\bfk}^2(x)&=0,
\\
[-\p_x^2+\bar{M}_{22}^2(x)]f_{\bfk}^2(x)+\bar{M}_{21}^2(x)f_{\bfk}^1(x)&=0.
\eal 
\end{subequations}
For a quantum state with a vanishing particle number on top of the mean field, we can express the two-point functions in terms of the mode functions as follows
\bal 
\label{eq:mode-2point}
G_j=G_j(x,x)=\int\widetilde{\d k} |f_{\bfk}^j(x)|^2,~~~~
K(x,x)=\bar{K}(x,x)=0.
\eal 
Therefore, an alternative approach to solve the quantum system is to evolve the mode functions via \eqref{eq:mode-eom}, compute $G_i$, and then evolve the mean field $\Phi_i$ via \eqref{eq:1point-eom}. Because there are as many mode functions as Fourier modes, this is numerically expensive, and so we further approximate the system by ensembles of field configurations as outlined in the next section.

\subsection{Stochastic ensemble average}
\label{sec:ensembleaverage}

Given that the mode functions satisfy linear equations of motion, we can employ a statistical method to approximate the quantum system \cite{Borsanyi:2007wm}. Suppose that we have a stochastic field $\varphi_j^e(x)$ that is expressed in terms of the mode functions above as follows,
\be 
\label{eq:pert-ensemble}
\cph_j^e(x)=\int\widetilde{\d k}\[c_{\bfk}^{j,e}f_{\bfk}^j(x)+c_{\bfk}^{j,e*}f_{\bfk}^{j*}(x)\],
\ee 
where the random variable $c_{\bfk}^{j,e}$ are drawn from a normal distribution with zero mean and has the following variance 
\be 
\label{eq:particleNumber-ensemble}
\lan c_{\bfk}^{j,e*}c_{\bfl}^{j,e}\ran_E=(2\pi)^{d}\oi_k\di^{d}(\bfk-\bfl)
\ee 
with $\lan\ran_E$ denoting the ensemble average. It is easy to show that $\varphi_j^e(x)$ satisfies the following equations
\begin{subequations}
\label{eq:pert-eom}
\bal
[-\p_x^2+\bar{M}_{11}^2(x)]\cph^e_1(x)+\bar{M}_{12}^2(x)\cph^e_2(x)&=0,
\\
[-\p_x^2+\bar{M}_{22}^2(x)]\cph^e_2(x)+\bar{M}_{21}^2(x)\cph^e_1(x)&=0.
\eal 
\end{subequations}
Now, the key observation is that the ensemble two-point functions of $\varphi_j^e(x)$ are numerically the same as the quantum two-point functions of $\varphi_j(x)$,
\begin{subequations}
\label{eq:point-ensemble}
\bal 
\lan\cph_1^e(x)^2\ran_E-\lan\cph_1^e(x)\ran_E^2&=G_1,
\\
\lan\cph_2^e(x)^2\ran_E-\lan\cph_2^e(x)\ran_E^2&=G_2,
\\
\lan\cph_1^e(x)\cph_2^e(x)\ran_E-\lan\cph_1^e(x)\ran_E\lan\cph_2^e(x)\ran_E&=K=\bar{K}.
\eal 
\end{subequations}
That is, the quantum averages over operators can be replaced by the ensemble averages over some auxiliary stochastic fields. To recapitulate, instead of directly solving the one- and two-point functions, we now solve the equations of motion for the one-point functions \eqref{eq:1point-eom}, together with running a number of realizations of the equations of motion for the auxiliary stochastic fields \eqref{eq:pert-eom} and taking the ensemble averages to get the two-point functions that feed into \eqref{eq:1point-eom}. For a $d$ dimensional spatial lattice, if the number of realizations needed to get satisfactory statistical convergence is much smaller than the number of the lattice points in the $d$ dimensional spatial lattice, ($N^d\times \mc{E}\ll N^d\times N^d$) we gain computational speed-up. As we will see, for the present case the speed-up is rather significant.  

\subsection{Renormalization}
\label{sec:renorm}

When evaluated at the same spacetime point, the two-point functions are formally divergent in the continuum limit. The lattice regularization turns the divergences into finite, but potentially large contributions to the mass and couplings. While we can run the lattice simulations with the bare quantities, to connect with physical results, or to compare with the classical results, we need to renormalize our input parameters.  

While we start our simulations such that the mean field is the classical Q-ball solution, the perturbative field $\varphi_i$ is set to be initially in the vacuum,
\be 
\label{eq:mode-initial}
f_{\bfk}^j(x)=e^{ikx} , \quad \p_tf_{\bfk}^j(x)=-i\oi_ke^{ikx}.
\ee 
With this condition we have 
\be 
\label{eq:2point-initial}
G^{(\rm init)}_j(x,x) =A=\int\f{\d^d k}{(2\pi)^{d}2\oi_k}.
\ee 
This quantity contains quadratic and logarithmic divergences in 3+1D and a linear divergence in 2+1D. What this means is that if the renormalised physical parameters (the ones used in a purely classical simulation) are to be $m_r$, $\li_r$, $g_r$, we need to simulate the quantum system with bare parameters defined as
\bal 
\label{eq:renomarlization}
&\li_r=\li-12gA,
\\
&m_r^2=m^2-4\li A+24gA^2,
\\\label{eq:lambdar}
&g_r=g.
\eal  
Having found $A$ from (\ref{eq:2point-initial}), these equations are solved sequentially, starting with $g$, then $\lambda$, then $m$. We will insert the classical Q-ball as the initial conditions. This is not perfectly matched with a space-independent vacuum state for the modes, and so there will be an initial adjustment and relaxation as the modes discover the Q-ball background. This is a small effect which could in principle be reduced further by gradually (adiabatically) dialing in the Q-ball. 

\subsection{Numerical setup}
\label{sec:setup}

In classical numerical simulations, it can be useful to adopt the dimensionless variables
\be
\label{scaling10}
\t{x}^{\mu}=m_rx^{\mu},~~~\t{\cph}=\cph\li_r^{1/2}/m_r,~~~\t{g}=g_rm_r^2/\li_r^2 ,
\ee
with which the action becomes 
\be 
\label{eq:action-program}
S=\f{m_r^{4-D}}{\li_r}\int \d^D x \(-\Big|\f{\p\t{\cph}}{\p\t{x}^{\mu}}\Big|^2-|\t{\cph}|^2+|\t{\cph}|^4-\f43\t{g}|\t{\cph}|^6\).
\ee 
We see that the only free parameter in this action is now $\t{g}$. In this paper, the fiducial choice for $\t{g}$ is $\t{g}=3/8$ allowing us to directly compare with previous classical results in \cite{Battye:2000qj, Xie:2021glp}. With this choice of $\t{g}$, the range where the classical Q-ball exists is $1/\sqrt2\leq\t{\oi}\equiv \omega/m_r<1$.

In the quantum theory, however, such a rescaling of the field also implies a normalization of the quantum (vacuum) modes. Hence, while classically eliminating the $m_r$ and $\li_r$ dependence in the equations of motion is just a reparameterization, in the quantum theory it amounts to specific choices for those parameters, if we want to leave the quantum vacuum unchanged. Therefore, we instead adopt the following dimensionless variables (in 2+1D)
\be 
\t{x}^{\mu}=m_rx^{\mu},~~~\t{\cph}=\cph/m_r^{1/2},~~~\t{\li}=\li_r/m_r,~~~\t{g}=g_r,
\ee 
where all quantities are cast in units of $m_r$. The action now depends on two free parameters 
\be 
\label{quantum_action1}
S=\int \d^D x \(-\Big|\f{\p\t{\cph}}{\p\t{x}^{\mu}}\Big|^2-|\t{\cph}|^2+\t{\li}|\t{\cph}|^4-\f43\t{g}|\t{\cph}|^6\).
\ee 

This allows us to explicitly investigate how the classical limit arises in the limit of large (in field amplitude) Q-balls. To illustrate the point, we briefly consider a single scalar field with the potential $V(\Phi)=m_r^2\Phi^2/2-\lambda_r \Phi^4/4+g_r\Phi^6/6$. The equation of motion in the Hartree approximation is then
\bal 
[-\p_x^2
+m_r^2-\li_r\Phi^2+g_r\Phi^4]\Phi= & -3\li_r(A-G)\Phi +10g_r(A-G)\Phi^3 
 -15g_r(A-G)^2\Phi ,
\eal 
where $G$ is the correlator (\ref{eq:mode-2point}) but for the one real field and $A$ is defined in \eqref{eq:2point-initial}. We can now construct a one-parameter family of models through the following rescaling of the parameters of the potential and the field amplitude (but not the quantum modes)
\be
\label{aidef1}
\t{\li}\to \ai\t{\li},~~\t{g}\to \ai^2\t{g},~~\t{\cph}\to\t{\cph}/\sqrt{\ai},
\ee
where we have in mind that $\alpha<1$. It is straightforward to find, that the net result is to replace the equation of motion by
\bal 
[-\p_x^2
+m_r^2-\li_r\Phi^2+g_r\Phi^4]\Phi= & -3\li_r\ai(A-G)\Phi +10g_r\ai(A-G)\Phi^3
 -15g_r\ai^2(A-G)^2\Phi . 
\label{eq:rescaledmodes}
\eal 
Since the equation of motion is unchanged on the left-hand side, we know that there is a Q-ball solution with the same properties as the original one, but scaled up by $1/\sqrt{\alpha}$. And that in the quantum theory, evolving that scaled-up Q-ball with the original quantum corrections on the right-hand side, is equivalent to evolving the original Q-ball with quantum corrections scaled down by $A-G\rightarrow \alpha (A-G)$ \footnote{Since the amplitude of the quantum modes is set by $\hbar$, this may be thought of as dialing down $\hbar$, thereby achieving the classical limit. In fact, it is a scaling up of the Q-ball field amplitude to larger and larger occupation numbers, which is indeed the standard classical limit. We note that large occupation numbers does not necessarily imply large charge $Q$.}. In this way, we see that for any Q-ball in this potential, we can straightforwardly generate larger and larger (in the sense of field amplitude) Q-balls, for which quantum corrections are of less and less importance. For most of our simulations, we retain the fiducial model $\t{\li}=1,\t{g}=3/8$.

\begin{figure}[tbp]
\centering
\includegraphics[width=.45\textwidth]{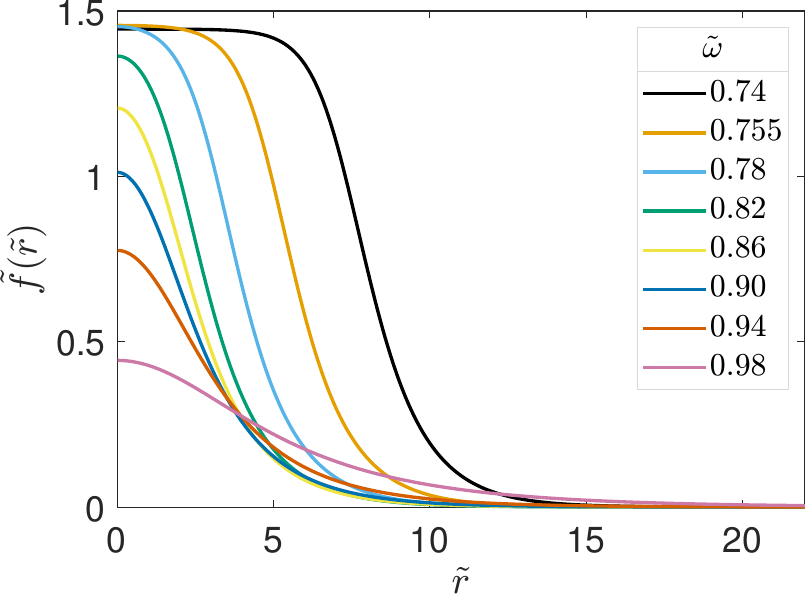}
\hfill 
\includegraphics[width=.49\textwidth]{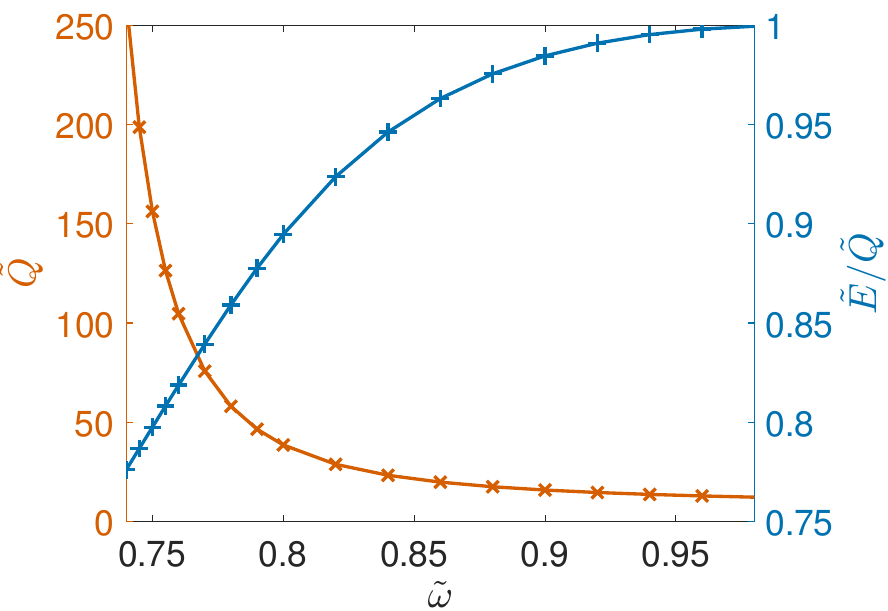}
\caption{\label{fig:qball2D}
Q-ball profiles for various internal frequencies (left) and dependence of charge and energy-to-charge ratio on the internal frequency (right). 
}
\end{figure}

\begin{figure}[tbp]
\centering
\includegraphics[width=.45\textwidth]{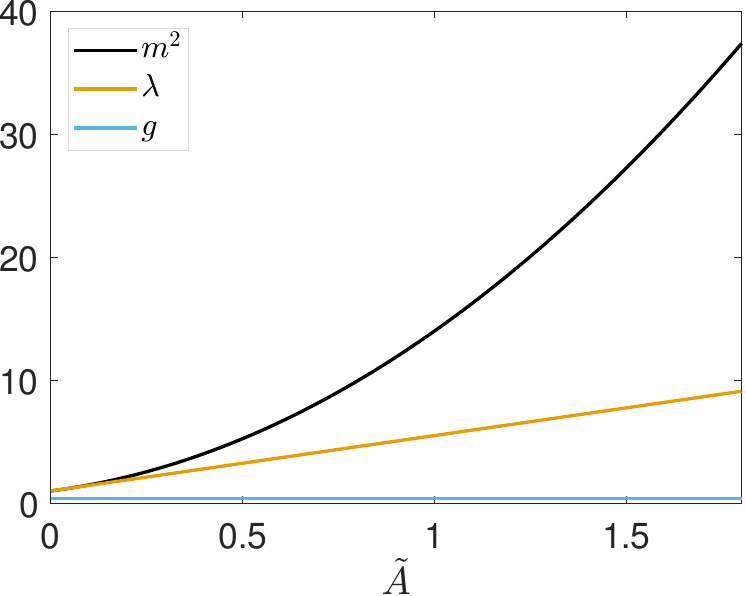}
\hfill 
\includegraphics[width=.49\textwidth]{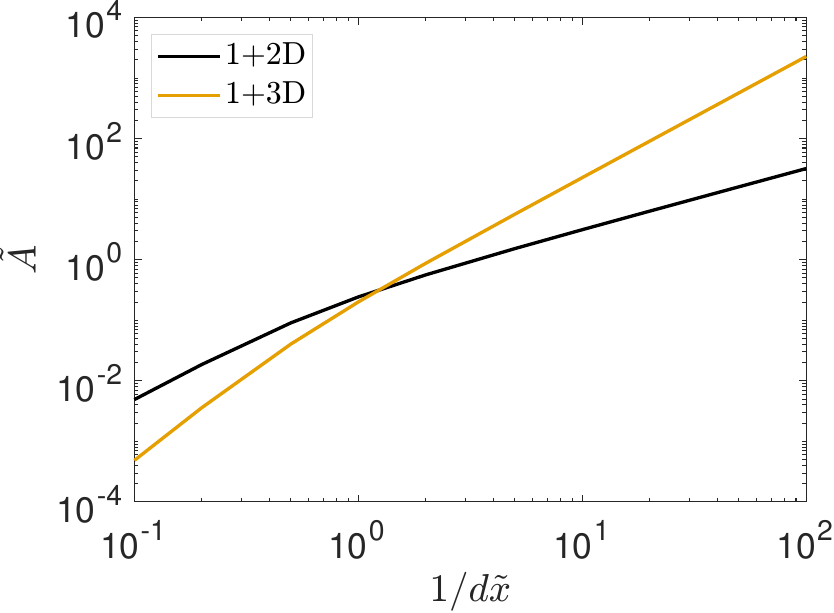}
\caption{\label{fig:renormalization}
Running of bare couplings $m^2,\li,g$ with $A$ (left) and running of $A$ with lattice spacing $dx$ (right). 
}
\end{figure}

For simplicity, we focus on 2+1D Q-balls in the numerical simulations. The 2+1D Q-ball profile functions are shown for different internal frequencies in the left plot of Figure \ref{fig:qball2D}, and the charge and charge to energy ratio of the corresponding Q-balls are shown in the right plot of Figure \ref{fig:qball2D}. The Q-balls with small frequencies are known as thin-wall Q-balls while those with large frequencies are known as thick-wall Q-balls, due to the thicknesses of their surfaces, shown in Figure \ref{fig:qball2D}. 
The numerical lattice we use has a spacing of $m_rdx=0.4$, which sets the largest momentum to be $k_{\text{max}}/m_r=\sqrt{d}\pi/(m_rdx)\sim O(10)$. The UV cutoff of the lattice is thus one order of magnitude greater than $m_r$. In Figure \ref{fig:renormalization}, we show on the left the running of the bare couplings with $A$, and on the right how $A$ runs with $dx$. ($A$ in principle also depends on the size of the lattice, but the integral is clearly dominated by the part near the UV cutoff.) In the 3+1D case at $m_rdx=0.5$, the value of $A$ from our code is consistent with that of \cite{Tranberg:2013cka}. Other numerical parameters will be mentioned as we go along when presenting the relevant numerical results.
Our code uses the {\tt LATfield2} package \cite{Daverio:2015ryl} to implement parallelization. It uses a second order finite difference stencil for the spatial derivatives and evolves in time with the leapfrog method. The spatial boundaries are periodic. When necessary, we use a sufficiently large lattice to alleviate the impact from the returning waves due to the periodic boundaries. 

To examine how the quantum corrections differ from the classical dynamics, we mainly monitor the Noether charge and the correlation functions. For the expectation value of the charge operator, since the cross terms vanish, it is natural to split it into a mean field part $Q_{\Phi}$ and a perturbation part $Q_G$
\bal 
\label{eq:charge-quantum}
&Q=\lan \hat Q \ran=Q_{\Phi}(t)+Q_G(t)
\eal
with 
\bal
\label{QPhi}
Q_{\Phi}(t)&=\int \d^d x(\Phi_2\p_t\Phi_1-\Phi_1\p_t\Phi_2),
\\
\label{QG}
Q_G(t)&=\int \d^d x\lan\cph_2\p_t\cph_1-\cph_1\p_t\cph_2\ran.
\eal 
The quantum average of operators $\cph_j$ in \eref{QG} can further be replaced by the ensemble average of stochastic fields $\cph_j^e$. Similarly, we can also separate the charge density into the mean field part and the quantum perturbations part
\be
j^0 = j^0_\Phi + j^0_G .
\ee

\section{Single Q-ball}
\label{sec:singleQball}

\begin{figure}[tbp!]
\centering
\includegraphics[width=.5\textwidth]{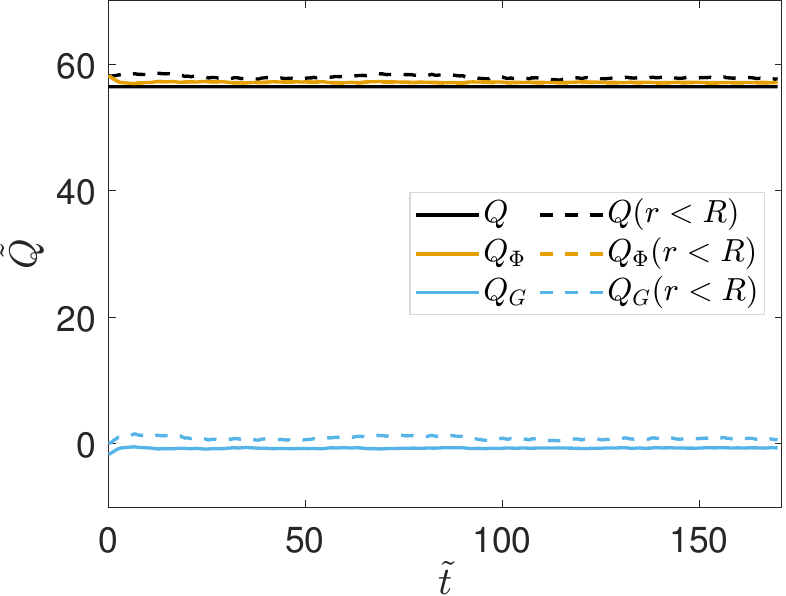}
\caption{\label{fig:charge-singleQball_1}
Quantum evolution of various (global) charges for a thin-wall Q-ball. The total charge over the whole simulation box is $Q=Q_{\Phi}+Q_G$, with $Q_{\Phi}$ being the mean field part and $Q_G$ being the ``quantum'' part, as defined in \eqref{eq:charge-quantum}, while $Q(r<R)=Q_{\Phi}(r<R)+Q_G(r<R)$ is the charge inside a disk of radius $\t{R}=17$ from the Q-ball center. $Q_{\Phi}$ is matched to the pure classical charge at $t=0$. The internal frequency of the Q-ball is $\t{\oi}=0.78$, and the Q-ball is centered at the origin.}
\end{figure}
In this section, we investigate the quantum dynamics of a single Q-ball. The Hartree ensemble approximation has been used to study the quantum stability of Q-balls previously in 3+1D \cite{Tranberg:2013cka}. Here we instead focus on a 2+1D Q-ball in detail, and we will see that some properties of Q-balls do depend on the spacetime dimensions, as is the case classically. Generically, we find that the inhomogeneous Hartree approximation leads to less corrections in 2+1D than in 3+1D. 

In Section \ref{sec:singleQball-classical}, we explore scenarios where the classical approximation closely aligns with the quantum corrected dynamics, while in Section \ref{sec:singleQball-quantum} we delve into scenarios where the quantum corrections significantly influence Q-ball dynamics. The former is when the relevant modes in the problem contain high occupation numbers such as in the case of thin-wall 
Q-balls, while the latter is exemplified in the case of thick-wall Q-balls. 

In this section, we generally set the number of lattice sites per direction to be $N=256$ and the number of the ensemble realizations to be $\mc{E}=20,000$. The time step is given by $d\t{t}/d\t{x}=0.05$, which is sufficient to ensure convergence. We have varied all these numerical parameters by $50\%$ and found no significant impact on the results. Empirically, an ensemble of size $\mc{E}=5,000$ is sufficient to capture statistical quantities. 

\subsection{Classical regime}
\label{sec:singleQball-classical}

\begin{figure}[tbp!]
\centering
\includegraphics[width=1\textwidth]{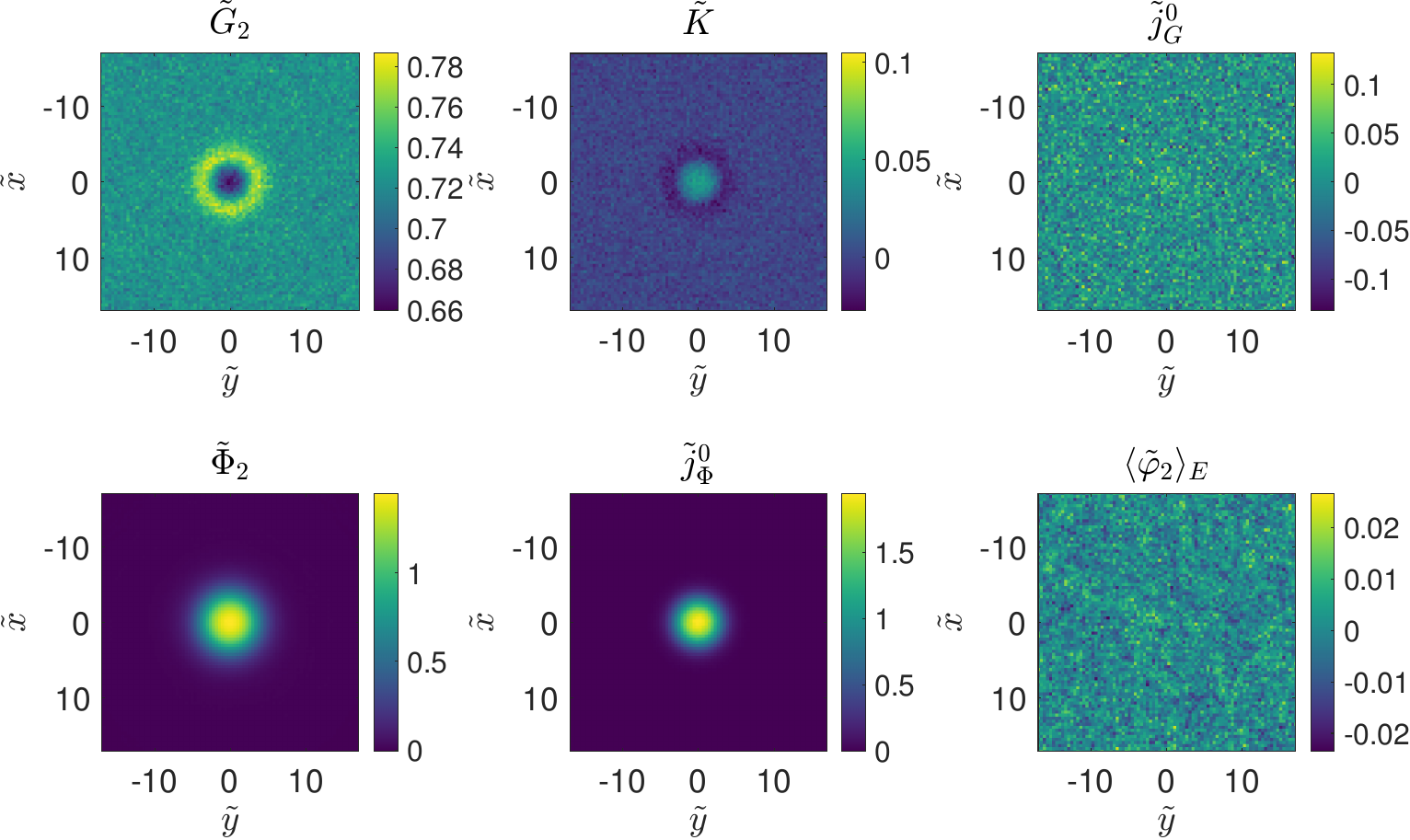}
\caption{\label{fig:field-singleQball_1}
Snapshots of various quantities at $\t{t}=80$, using the same setup as in Figure \ref{fig:charge-singleQball_1}. 
}
\end{figure}
One may expect that when the occupation numbers for all the relevant momentum modes are high, the classical field will be a good approximation for a given problem \cite{Aarts:1997kp,Aarts:2001yn}. For the case of a Q-ball, from Figure \ref{fig:qball2D}, we see that for a thin-wall Q-ball the charge density within the whole ball is very large, which indicates high occupation numbers for the relevant wave numbers, as reflected in its Fourier transform. We will see that the classical solution is indeed a good approximation for a thin-wall Q-ball.
\begin{figure}[tbp]
\centering
\includegraphics[width=.43\textwidth]{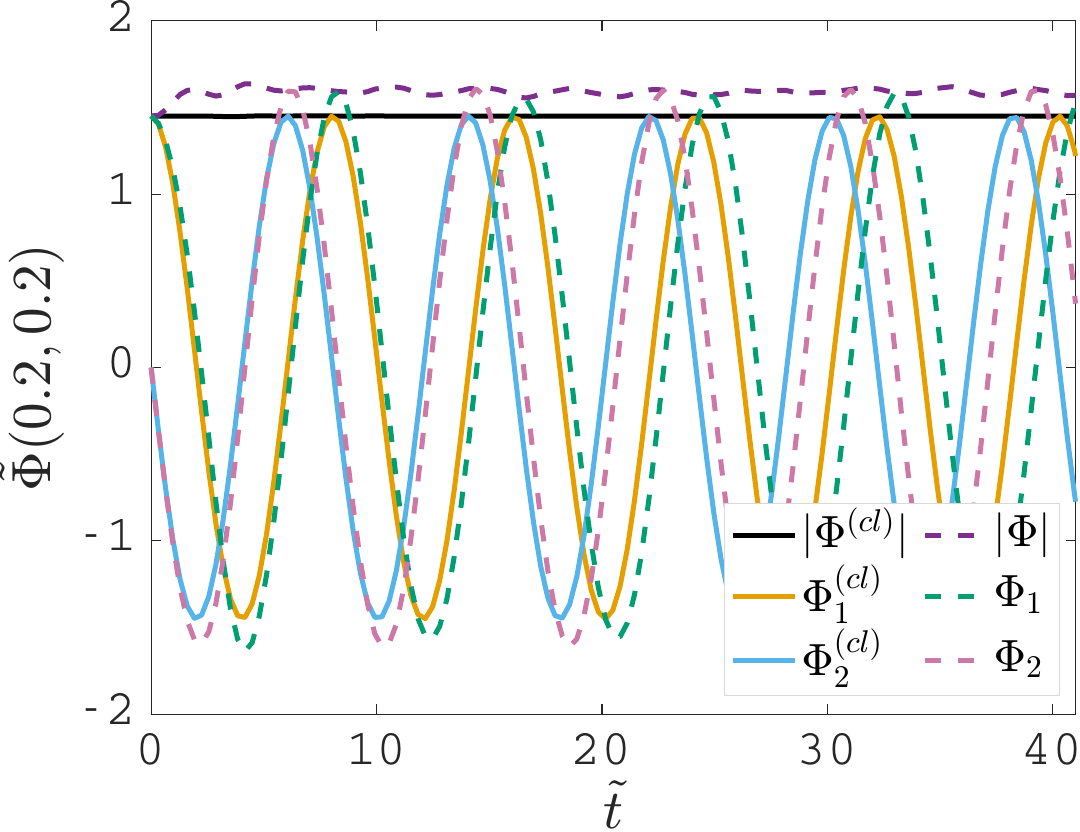}
\hfill
\includegraphics[width=.45\textwidth]{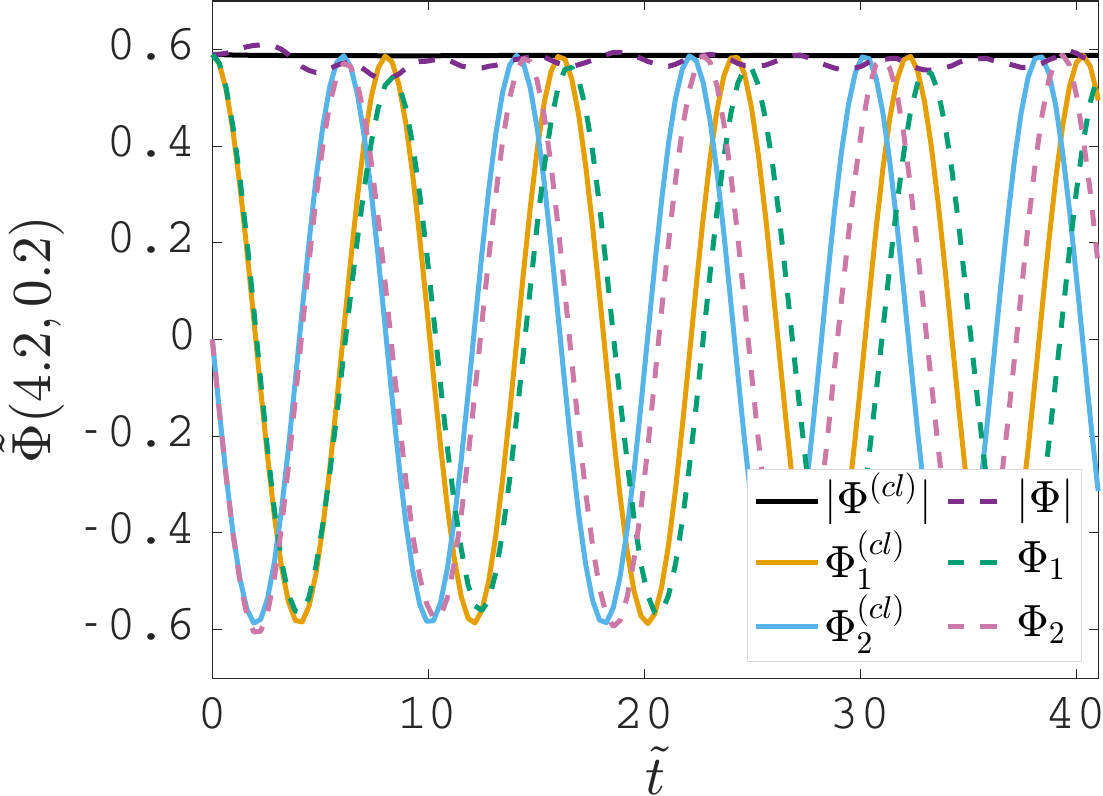}
\includegraphics[width=.45\textwidth]{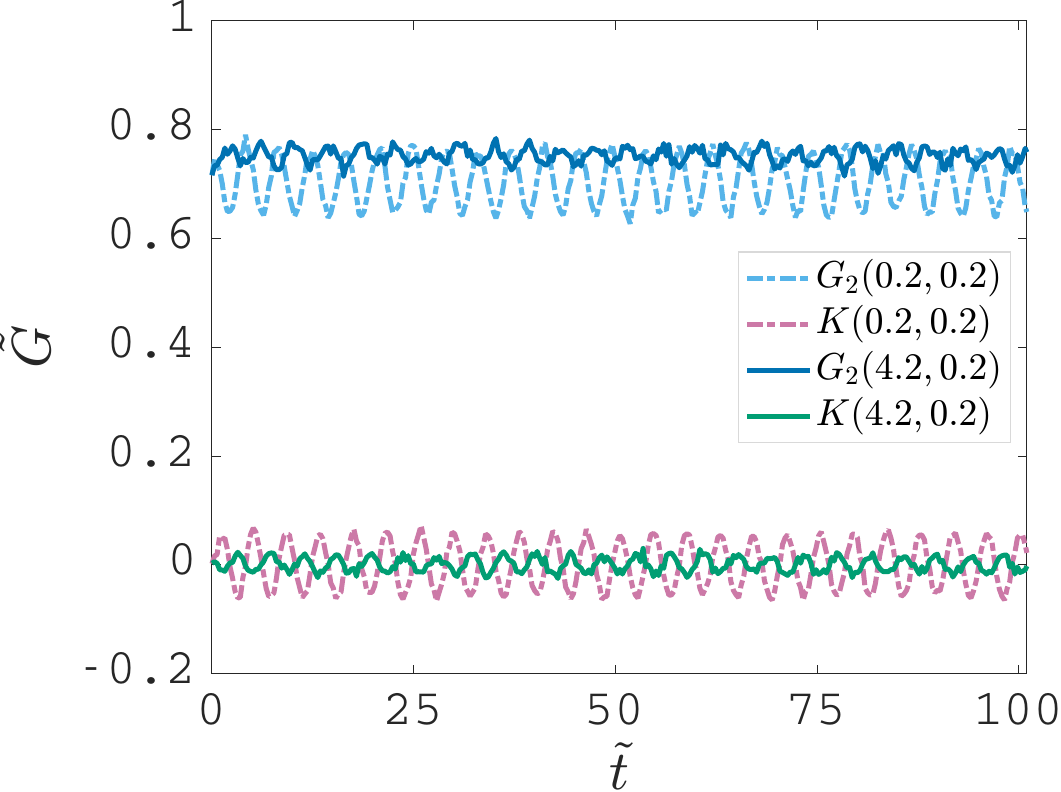}
\hfill
\includegraphics[width=.45\textwidth]{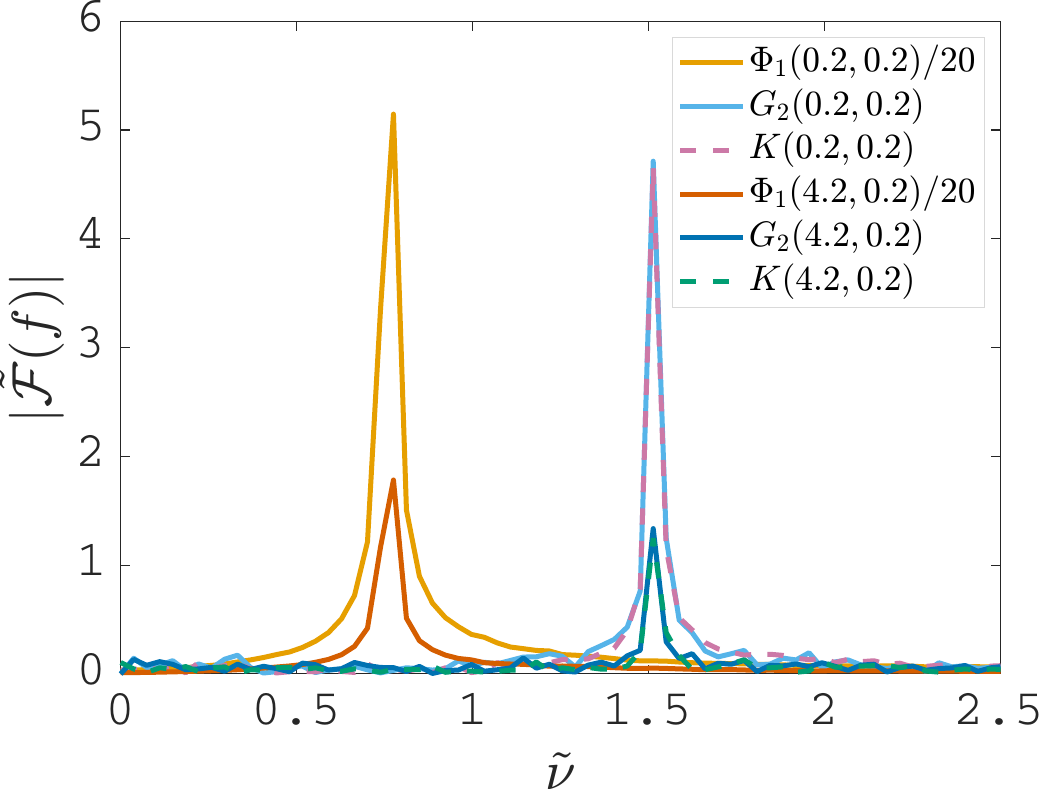}
\caption{\label{fig:nPointAtX-singleQball_1}
Evolution of one- and two-point functions at point $\t{\bfx}=(0.2,0.2)$ (left) and $\t{\bfx}=(4.2,0.2)$ (right). The corresponding classical solutions, labeled with ${}^{cl}$, are also plotted for comparison. The setup is the same as in Figure \ref{fig:charge-singleQball_1}. 
}
\end{figure}

Let us first examine some global properties of a quantum thin-wall Q-ball. In Figure \ref{fig:charge-singleQball_1}, the evolution of various charges are plotted, with the solid lines representing the charges in the entire lattice volume. As defined in \eref{eq:charge-quantum}, there are two sources for the total charge $Q=Q_{\Phi}+Q_G$, the charge from the mean fields $Q_{\Phi}$ and the charge from the quantum perturbations $Q_G$. For the thin-wall case, although both of $Q_{\Phi}$ and $Q_G$ oscillate slightly with time, the quantum perturbations are mostly un-excited, making $Q_{\Phi}$ a good approximation in this case. The conservation of the total charge $Q$ indicates that the simulation has achieved a good accuracy. $Q$ also approaches the charge of the corresponding classical Q-ball as the ensemble size approaches infinity, because initially $Q_G$ vanishes in this limit. To further verify that what is plotted really are the charges of the Q-ball, we also plot the charges inside a disk of a reasonably large radius $R$ around the Q-ball. Once again, we see that the quantum perturbations are effectively un-excited, and the quantum noise outside the Q-ball is under control.

Figure \ref{fig:field-singleQball_1} shows the snapshots of various fields at time $\t{t}=80$. First, we see that $K$ and $\lan\cph_2\ran_E$ vanish across the whole space, as expected. For the thin-wall Q-ball, we see that $G_2$ and $j^0_G$, which represent quantum effects and are rather homogeneous outside the Q-ball, are suppressed compared to the mean-field quantities such as $\Phi_2$ and $j^0_\Phi$. The behavior of the mean field $\Phi_2$ closely resembles that of the classical Q-ball.

Next, let us examine some local behaviour within the quantum Q-ball. In Figure \ref{fig:nPointAtX-singleQball_1}, we present the evolution of one- and two-point functions at two distinct points: one near the Q-ball center and another near its surface. At both points, the one-point functions $\Phi_j$ closely resemble the corresponding classical fields. We also plot the absolute value of the complex field $|\Phi|=|\Phi_1+i\Phi_2|$, and find that the absolute value $|\Phi|$, which corresponds to the profile function in the classical case, remains constant during the evolution. In our simulations, $G_{1}$ and $G_2$ oscillate slightly, so does $K$. 

The bottom right plot of Figure \ref{fig:nPointAtX-singleQball_1} displays the Fourier transforms of the mean fields' and the correlators' evolution for the two points. We can see that the dominant frequencies of $\Phi_j$ are $0.78 m_r$, essentially at the same frequency as the classical Q-ball, while the dominant frequencies of $G_j$ and $K$ are twice that. The Fourier transforms of the mean fields on the Q-ball surface are similar but with less power in the peak frequency. 

\subsection{Quantum regime}
\label{sec:singleQball-quantum}

For a smaller/lower density Q-ball that is not of the thin-wall type, we will see that the quantum effects are more prominent for the fiducial model. In this subsection, we look at how a Q-ball with $\t{\oi}=0.86$ evolves under the influence of quantum corrections.
\begin{figure}[tbp]
\centering
\includegraphics[width=.45\textwidth]{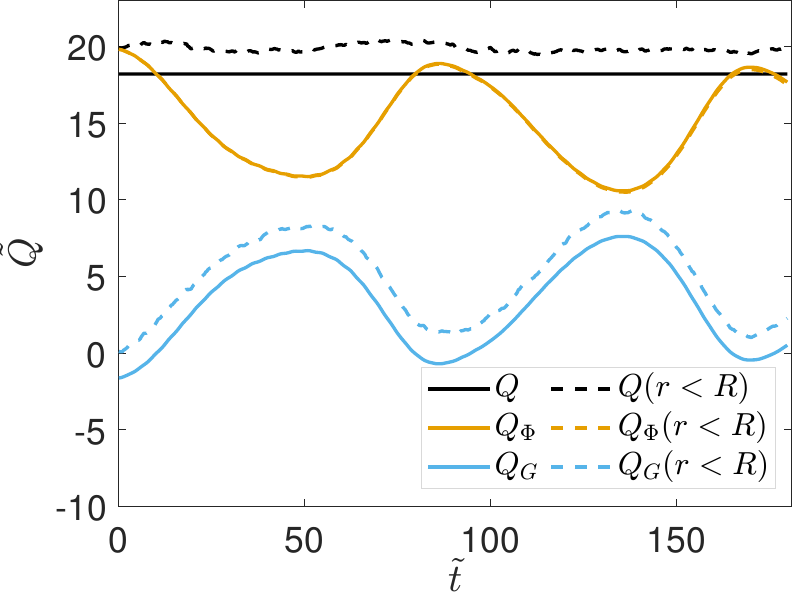}
\hfill 
\includegraphics[width=.45\textwidth]{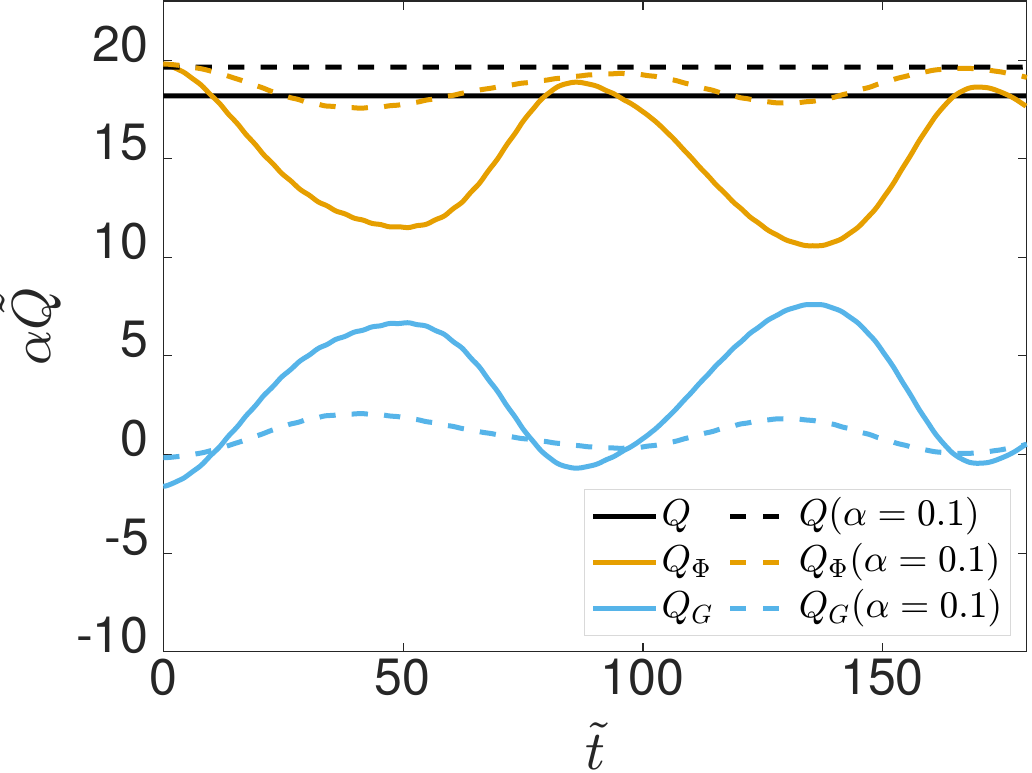}
\caption{\label{fig:charge-singleQball_2}
Quantum evolution of various charges for a thick-wall Q-ball, where $Q=Q_\Phi+Q_G$. The left plot is the fiducial case $\t{\li}=1,\t{g}=3/8$, and the right plot also shows the case $\t{\li}=0.1,\t{g}=3/800$. The internal frequency of the Q-ball is $\t{\oi}=0.86$ and we still choose $\t{R}=17$.
}
\end{figure}
\begin{figure}[tbp]
\centering
\includegraphics[width=.7\textwidth]{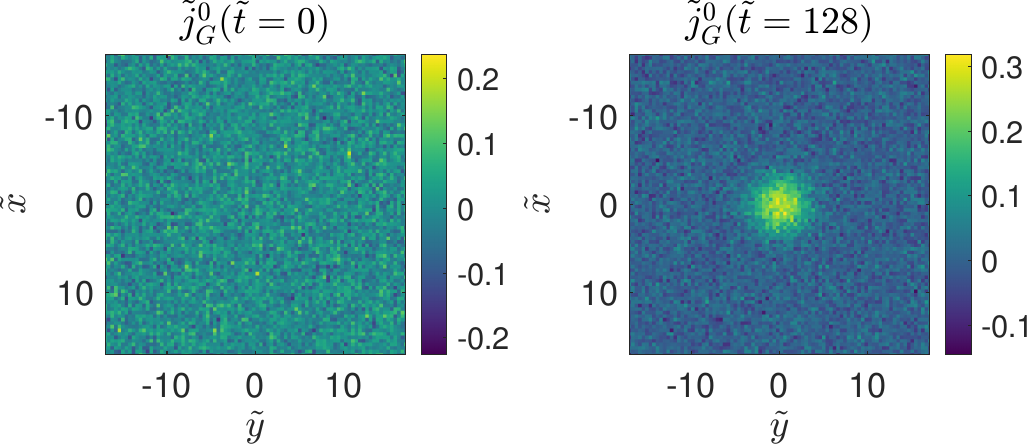}
\caption{\label{fig:samplej0G-singleQball_2}
Snapshots of $j^0_G$ at $\t{t}=0$ and $128$, using the same setup as in Figure \ref{fig:charge-singleQball_2}. 
}
\end{figure}
For such a Q-ball, a fascinating feature is that charges are exchanged between the mean fields and the perturbation fields, as shown in the left hand side of Figure \ref{fig:charge-singleQball_2}. A similar phenomenon has been observed previously in 3+1D \cite{Tranberg:2013cka}. At the initial time, almost all the charge is stored in the mean fields, up to some small ensemble fluctuations. During the time evolution, about half of the charge is exchanged between the mean fields and the quantum modes. When the Q-ball frequency $\oi$ becomes greater, the classical Q-ball will become smaller, and the mean field and mode charges will swap with a greater frequency. Even though the classical approximation breaks down in this case, most of the charge still condenses to a localized region, exhibiting the existence of a quantum version of the soliton, as shown by the curves with $r<R$. 

The right-hand side of Figure \ref{fig:charge-singleQball_2} shows the evolution following a rescaling of the form (\ref{aidef1}), with $\alpha=0.1$. For comparison, we in practice display a simulation with a scaled down $G$ as in (\ref{eq:rescaledmodes}). We see that indeed for this larger Q-ball, the effect of quantum interactions is much smaller. We may continue to do this (see also Figure \ref{fig:csq-diffLambda}), until the Q-ball is so large that quantum effects are negligible and classical simulations are exact. Figure \ref{fig:samplej0G-singleQball_2} shows the spatial distribution of the charge density $j^0_G$ field at the time $\t{t}=0$ and $128$. We see a localized charge density in the quantum modes, excited within the Q-ball.
\begin{figure}[tbp]
\centering
\includegraphics[width=.45\textwidth]{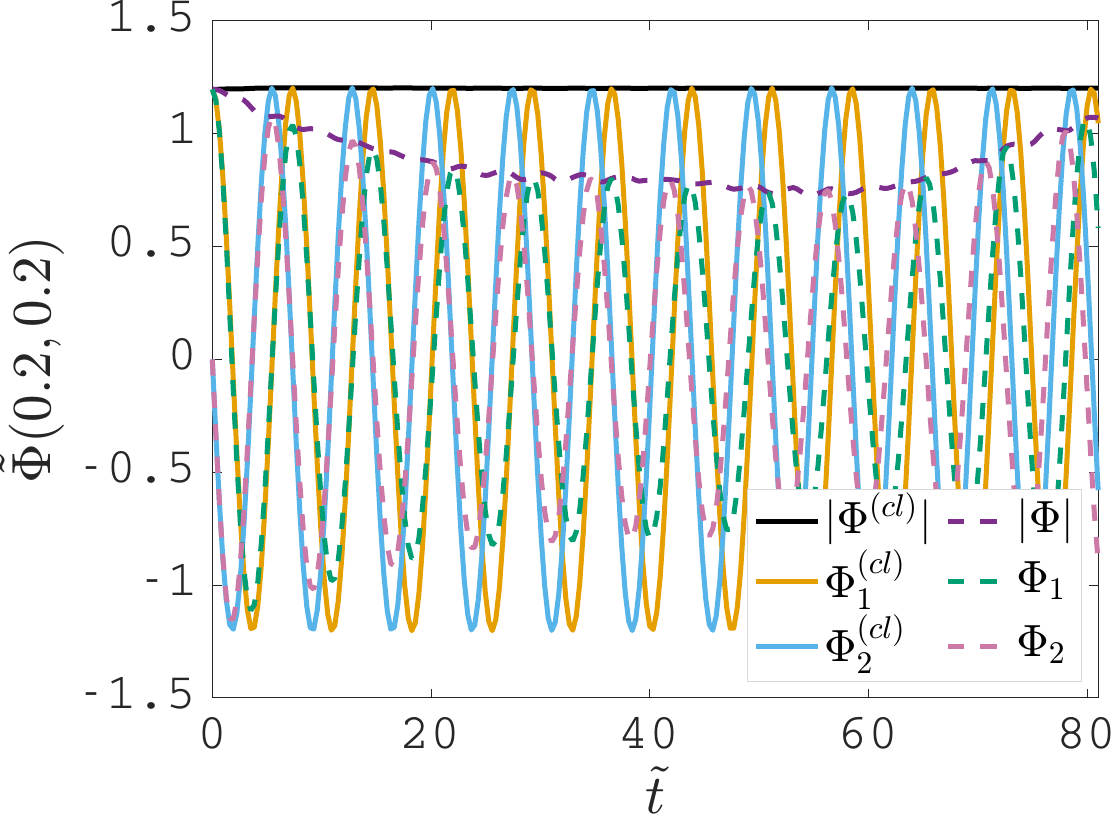}
\hfill
\includegraphics[width=.45\textwidth]{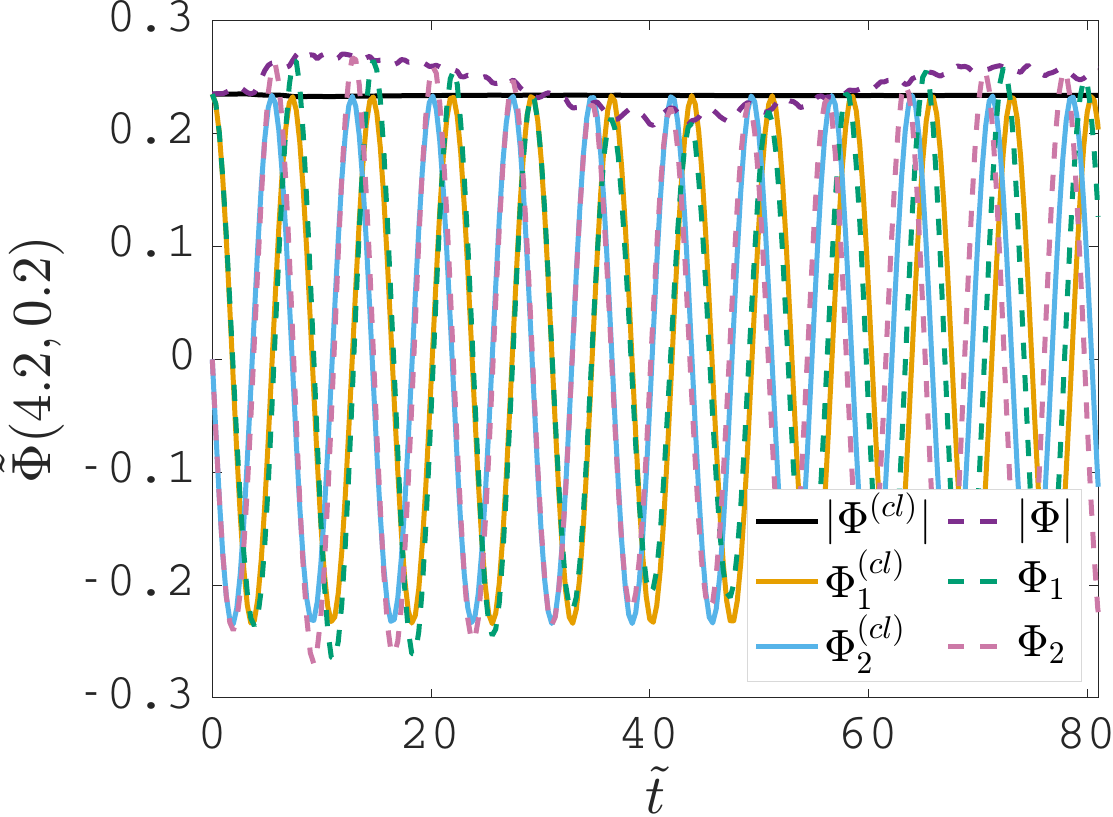}
\includegraphics[width=.45\textwidth]{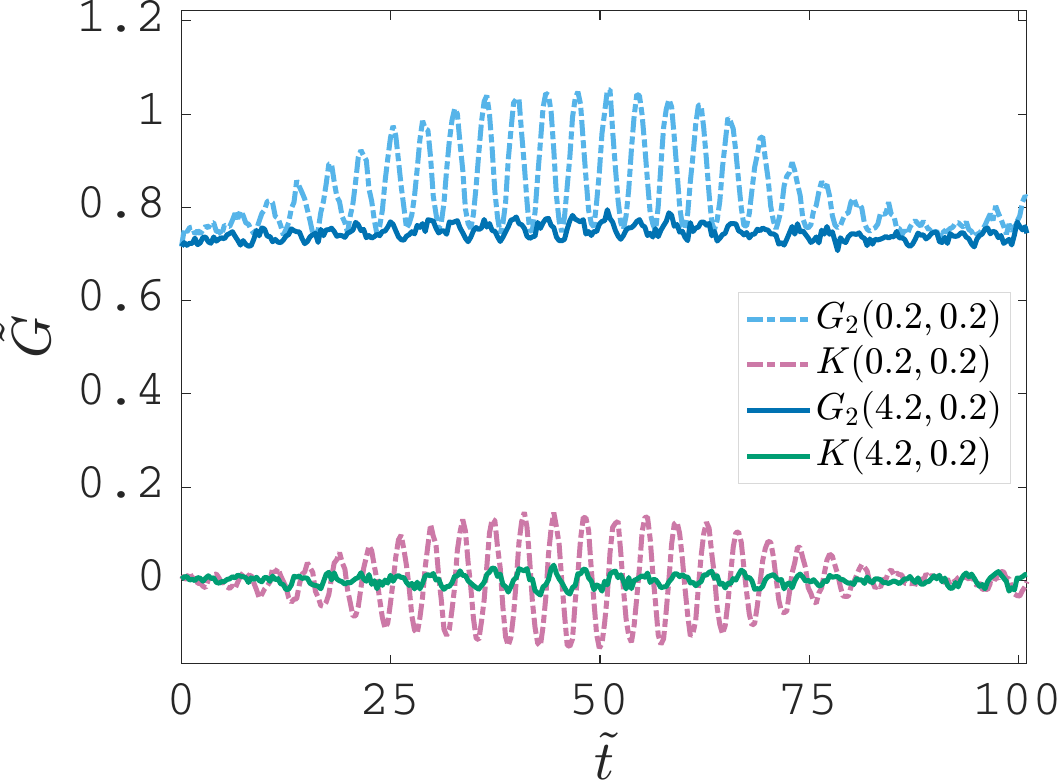}
\hfill
\includegraphics[width=.45\textwidth]{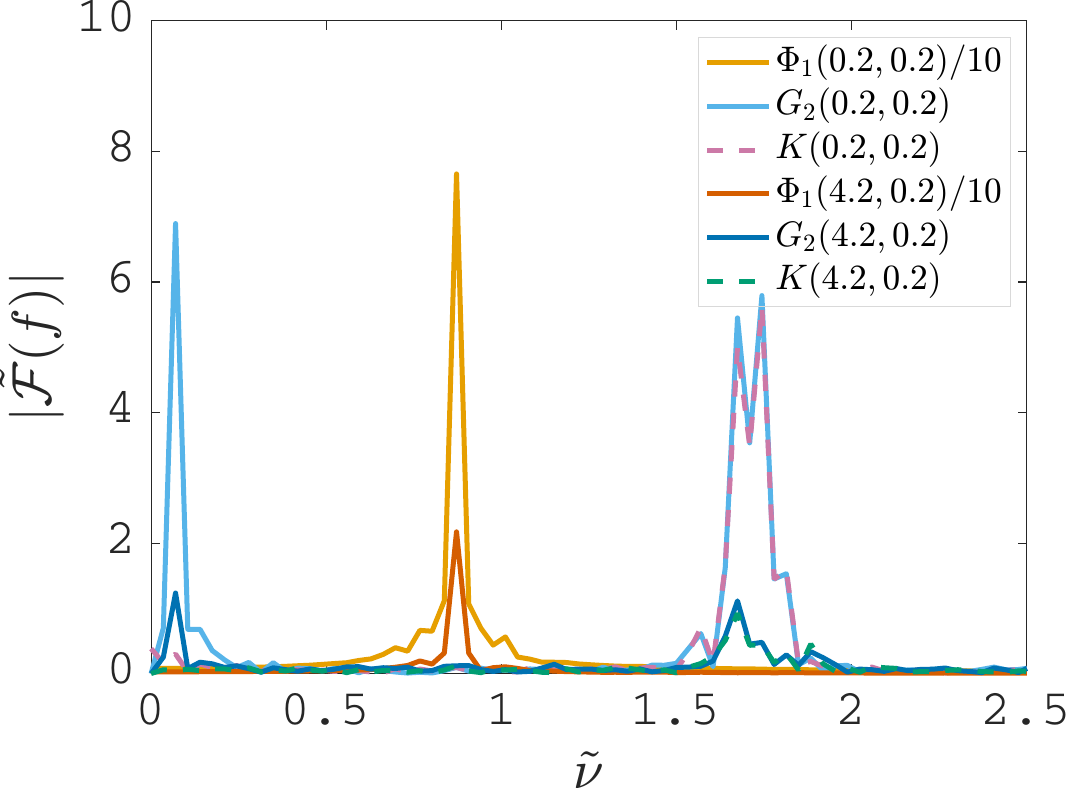}
\caption{\label{fig:nPointAtX-singleQball_2}
Evolution of one- and two-point functions at point $\t{\bfx}=(0.2,0.2)$ and $\t{\bfx}=(4.2,0.2)$. The corresponding classical solutions, labeled with $^{(cl)}$, are also plotted for comparison. The setup is the same as in Figure \ref{fig:charge-singleQball_2}. 
}
\end{figure}

In Figure \ref{fig:nPointAtX-singleQball_2}, we show the evolution of one- and two-point functions at the same points as in Figure \ref{fig:nPointAtX-singleQball_1}. While the frequencies and phases of the one-point functions still match those of the classical solution well, the amplitudes change periodically, especially at the points within the Q-ball. The changes in the one-point functions' amplitudes are matched by the modulations in the two-point functions, as they should.
\begin{figure}[tbp]
\centering
\includegraphics[width=.5\textwidth]{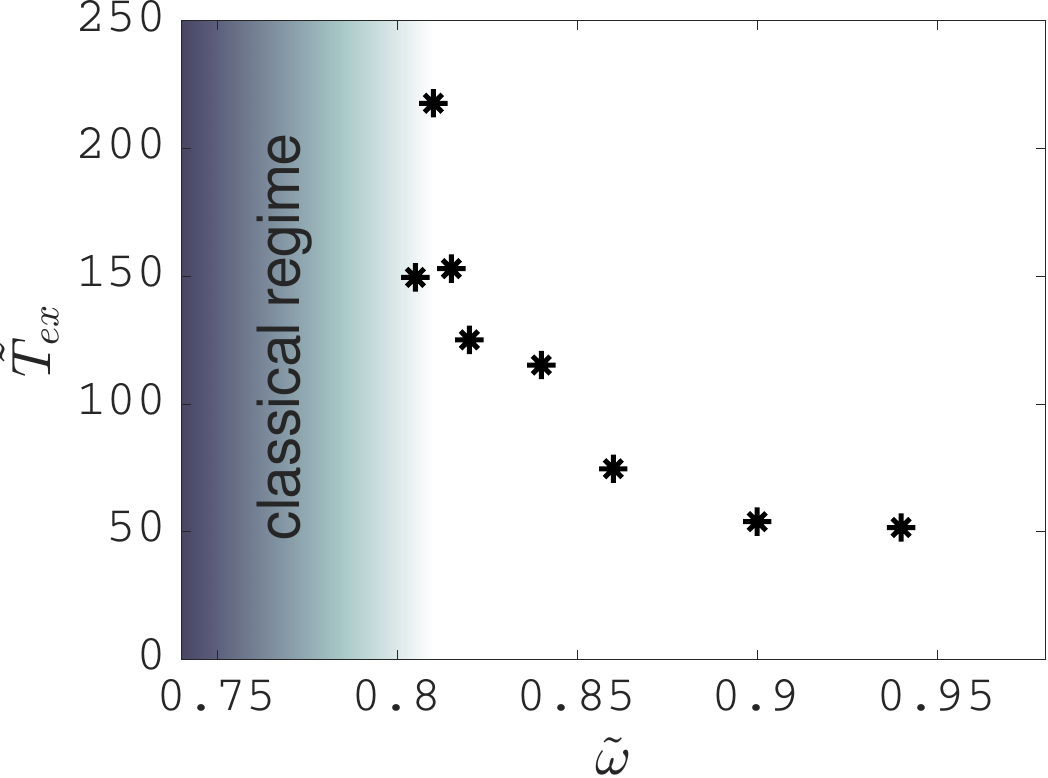}
\caption{\label{fig:exchangePeriod}
Average charge exchange periods between the mean fields and the perturbations with different Q-ball frequencies. The ensemble size $\mc{E}=5,000$ and $d\t{t}/d\t{x}=0.1$. 
}
\end{figure}
In Figure \ref{fig:exchangePeriod}, we see that the period of the charge exchange between the mean fields and the quantum modes, $T_{ex}$, increases as the frequency of Q-ball decreases, until the Q-ball enters the classical regime. Because the exchange period varies slightly with time, what is plotted in Figure \ref{fig:exchangePeriod} is the average period of the first dozen or so oscillations. Additionally, a greater proportion of the charge is exchanged as the Q-ball frequency decreases, at the frequency $\t{\oi}=0.81$, the whole charge is exchanged between the mean field and the quantum modes. 

In Ref \cite{Tranberg:2013cka}, it was shown that in 3+1D the Q-ball becomes unstable for a big portion of the frequency range where the classical Q-ball is stable. In contrast, we find that in 2+1D a classically stable Q-ball is usually also stable in the quantum theory, despite the charge exchange between the one- and two-point functions. However, we have only investigated a relatively small number of Q-ball frequencies. We leave a more comprehensive survey of the frequency space for future work.

\section{Interactions between multiple Q-balls}
\label{sec:multipleQball}

In this section, we investigate the interactions among multiple Q-balls. While there has been a lot of work studying the classical dynamics of such systems \cite{Axenides:1999hs,Battye:2000qj,Bowcock:2008dn,Brihaye:2009yr,Siemonsen:2023hko}, including quantum corrections to these scenarios has so far been left unexplored. 

\subsection{Well separated Q-balls}
\label{sec:phase}

Ref \cite{Battye:2000qj} has thoroughly investigated the interactions between two well-separated Q-balls with the same charge in the classical theory. If the two Q-balls with the same charge are in phase, they attract each other and coalesce into a larger Q-ball. If they are out of phase, they repel each other. In some cases, charge can be transferred from one Q-ball to the other. We will see that these findings are largely still valid with quantum corrections included. For simplicity, we will only present the results for the case of two Q-balls with phase difference $\Di\thi$ being $\pi/2$. 
\begin{figure}[tbp]
\centering
\includegraphics[width=.45\textwidth]{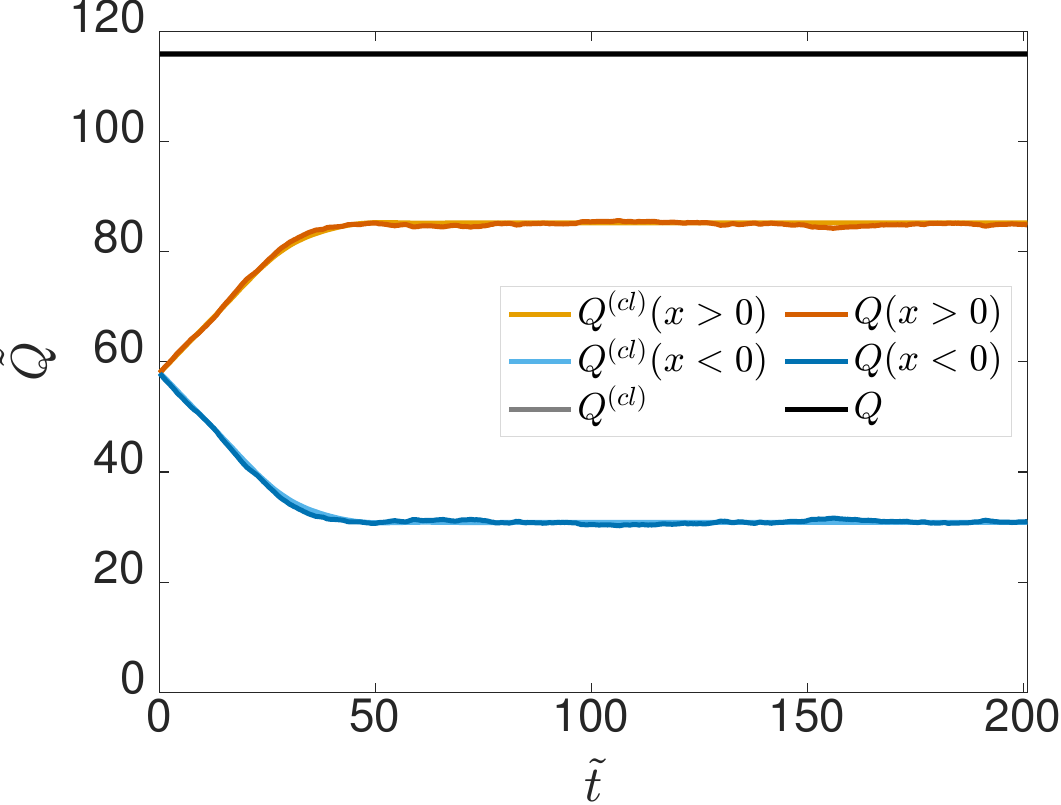}
\hfill
\includegraphics[width=.45\textwidth]{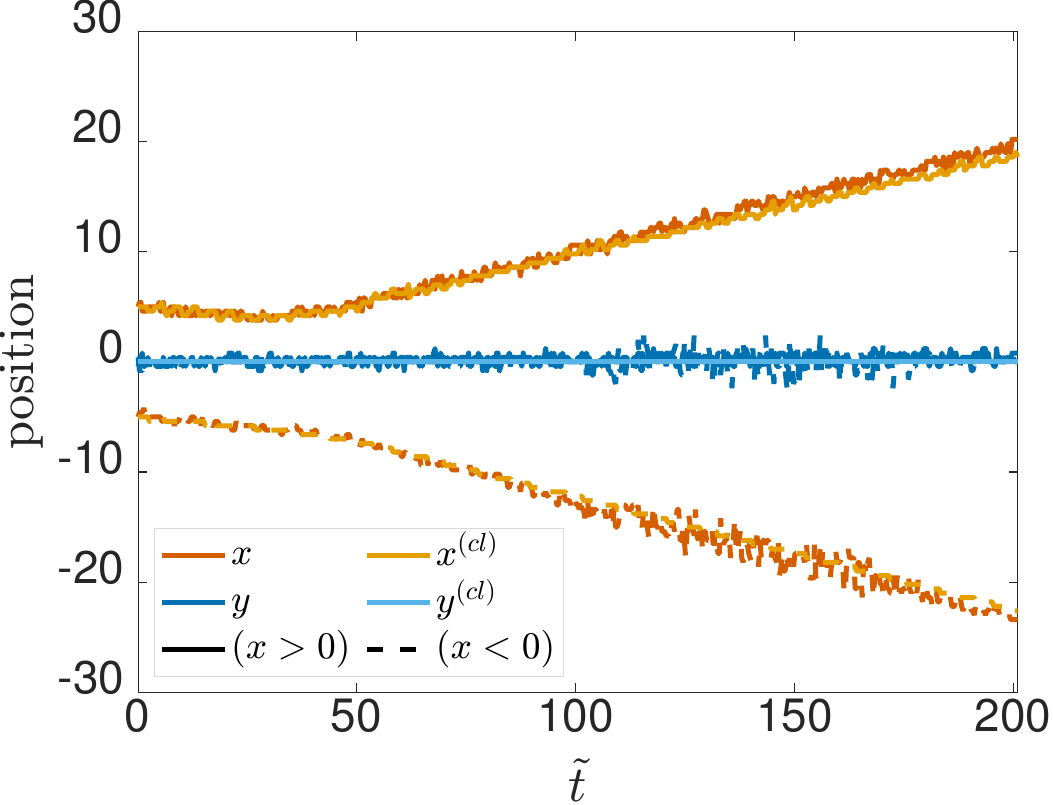}
\caption{\label{fig:comparison-twoQballs_wQ0p78}
The evolution of two well separated Q-balls with frequency $\t{\oi}=0.78$ and phase difference $\Di\thi=\pi/2$, placed at $(\pm5,0)/m_r$ initially. The two Q-balls repel each other. The left plot shows the evolution of the charges of the two Q-balls, and the right plot shows the evolution of their positions. ${}^{(cl)}$ denotes the corresponding classical simulation. The number of sites per dimension is $N=180$, the ensemble counts $\mc{E}=20,000$ realizations and $d\t{t}/d\t{x}=0.05$. 
}
\end{figure}
\begin{figure}[tbp]
\centering
\includegraphics[width=.5\textwidth]{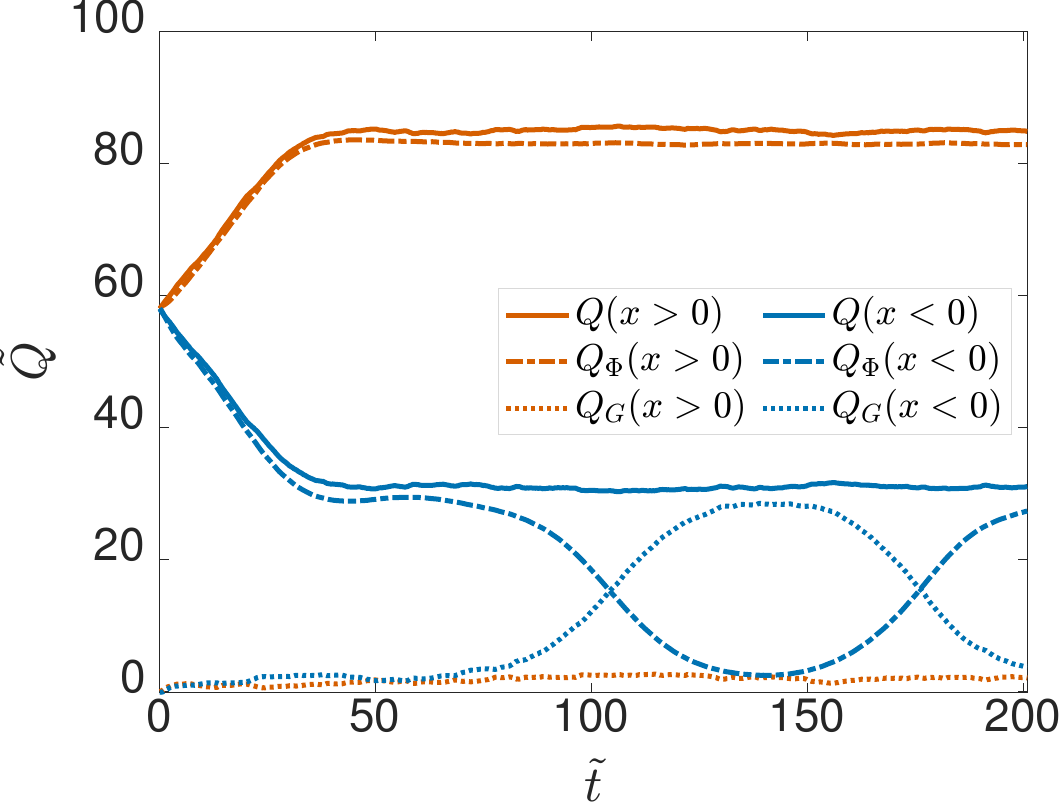}
\caption{\label{fig:charge-twoQballs_wQ0p78}
Various charges in different regions of the system. The setup is the same as Figure \ref{fig:comparison-twoQballs_wQ0p78}. After the initial charge transfer, the smaller Q-ball exhibits significant charge exchange between the mean fields and the quantum modes.
}
\end{figure}
Figure \ref{fig:comparison-twoQballs_wQ0p78} shows the evolution of two thin-wall Q-balls with phase difference $\pi/2$ placed at $(\pm5,0)/m_r$ initially. In both the classical and the quantum case, the two Q-balls repel each other. So the total charge in the $x>0$ region represents the charge of one Q-ball, and that in the $x<0$ region represents the other Q-ball. We can see that the Q-ball placed on the positive $x$-axis extracts charge from the one on the negative $x$-axis. The quantum effects do not affect the transfer rate in this process or the end states. The right plot shows the positions of the two Q-balls, which is obtained by locating the maxima of the charge density. 

In Section \ref{sec:singleQball}, we have seen that a Q-ball with sufficiently large charge is well approximated by the classical approximation and the charge in the quantum modes is negligible, while for a small Q-ball the charges in the mean fields and in the quantum modes are exchanged periodically. In Figure \ref{fig:charge-twoQballs_wQ0p78}, we see that the end state of the process in Figure \ref{fig:comparison-twoQballs_wQ0p78} is one large classical Q-ball and one small ``quantum'' Q-ball, which indeed starts to exchange its charges with the quantum modes.
\begin{figure}[tbp]
\centering
\includegraphics[width=.45\textwidth]{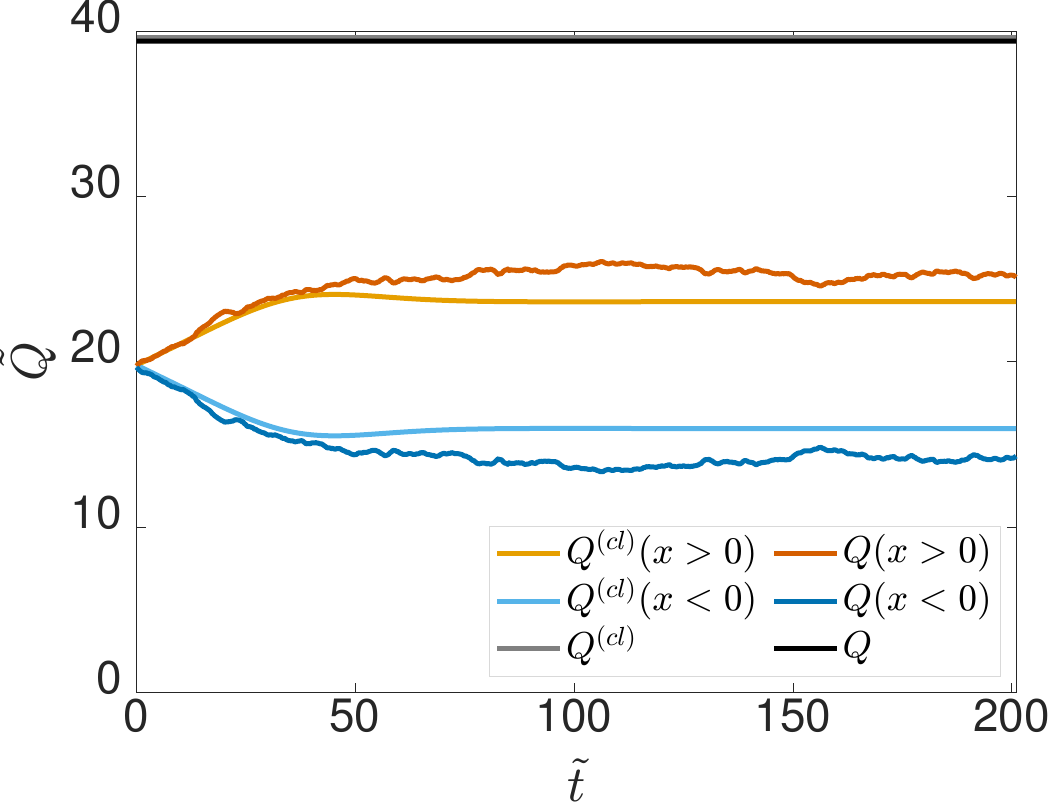}
\hfill
\includegraphics[width=.45\textwidth]{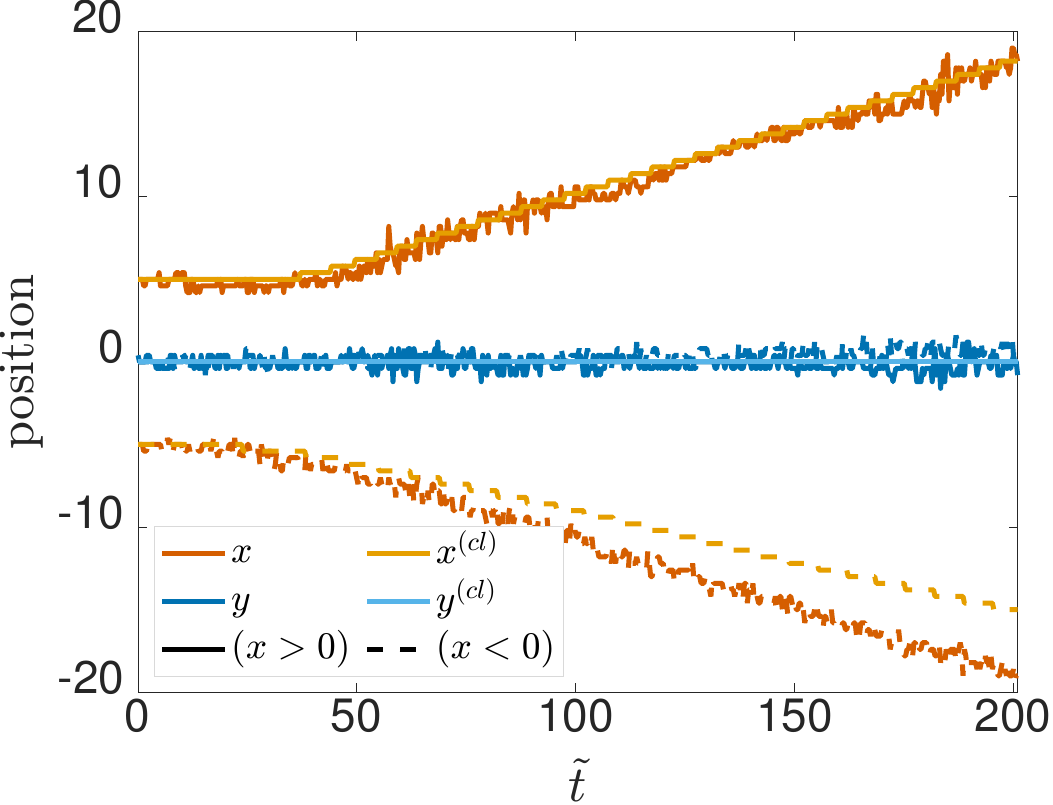}
\caption{\label{fig:comparison-twoQballs_wQ0p86}
The setup is similar to that in Figure \ref{fig:comparison-twoQballs_wQ0p78}, except that the initial Q-balls are smaller, with $\t{\oi}=0.86$. 
}
\end{figure}
\begin{figure}[tbp]
\centering
\includegraphics[width=.5\textwidth]{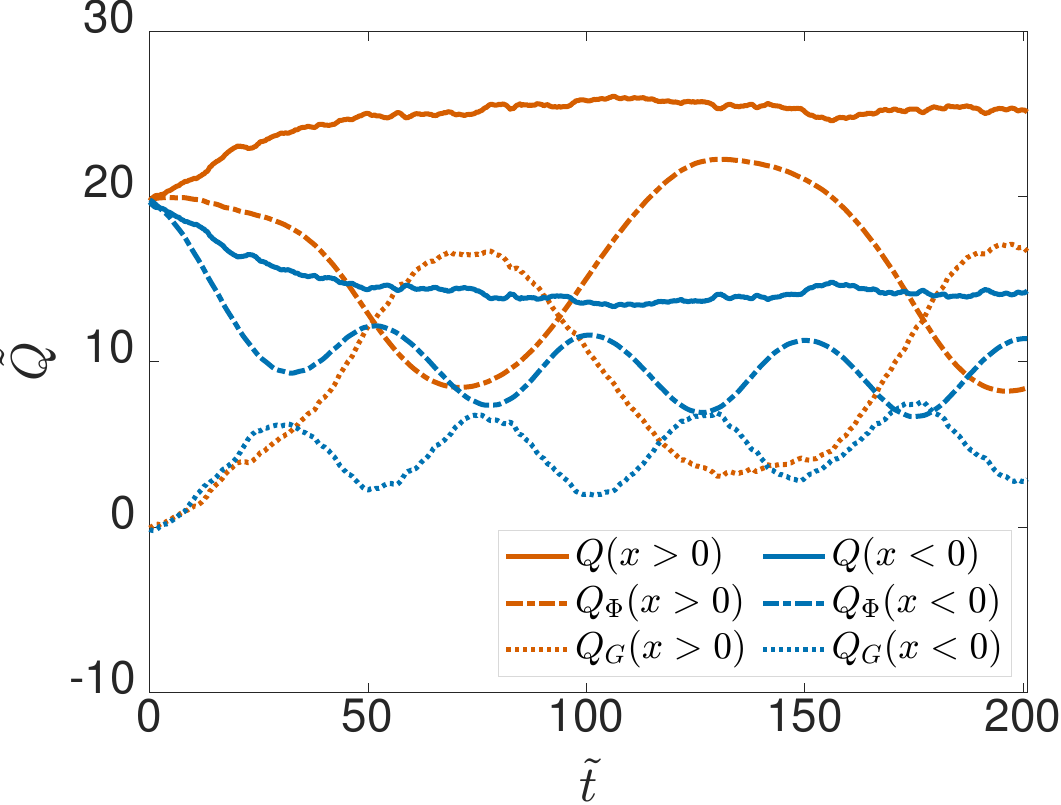}
\caption{\label{fig:charge-twoQballs_wQ0p86}
Various charges in different regions of the system. The setup is the same as in Figure \ref{fig:comparison-twoQballs_wQ0p86}. 
}
\end{figure}

Figure \ref{fig:comparison-twoQballs_wQ0p86} depicts the evolution of two small thick-wall Q-balls. In this case, differences are visible between the classical and quantum case. The Q-balls repel each other, and again some amount of charge is transferred from one Q-ball to the other. The quantum corrections do not affect the transfer rate, but they do affect the force between the two Q-balls. Particularly, the one that loses charge gets repelled further away, compared to that in the classical simulation, as can be seen in the right plot of Figure \ref{fig:comparison-twoQballs_wQ0p86}. In Figure \ref{fig:charge-twoQballs_wQ0p86}, we also separately plot the evolution of the mean field charge and the mode charge. We again find that although the total charge of the quantum evolution is close to the classical one, there is significant exchange of the charge between the mean field and the quantum modes. For small Q-balls, both the period of the charge exchange and the percentage of charges being exchanged are smaller.

\subsection{Collision of Q-balls}
\label{sec:collisions}

We now turn to the case where the Q-balls will attract and then collide. We consider small thick-wall Q-balls which, as we have seen previously, by themselves will only have charge exchange between the mean fields and the quantum modes. As we will see here, interestingly, the collision quenches the charge exchange between the mean fields and the perturbation fields. 

\begin{figure}[tbp]
\centering
\includegraphics[width=.5\textwidth]{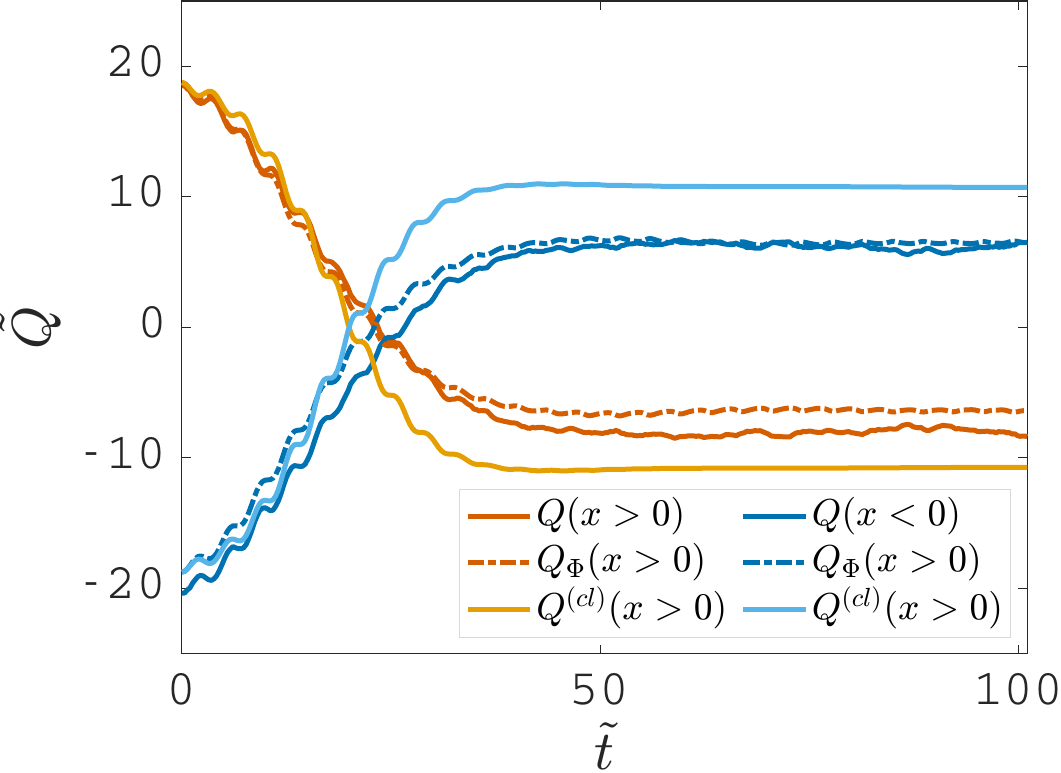}
\caption{\label{fig:swapOnce}
Two small Q-balls attract and collide. The periodic exchange of charge between the mean and perturbation fields are quenched by the collision. The two Q-balls, with $\t{\oi}=\pm0.86$, are placed at $(\pm3.4,0)/m_r$ initially. The classical and quantum case are plotted for comparison. The number of site per dimension is $N=256$, the number of ensemble realizations is $\mc{E}=10,000$ and $d\t{t}/d\t{x}=0.1$.
}
\end{figure}

In order to have the Q-balls automatically attracting each other, we prepare two thick-wall Q-balls with opposite charges. We place them on the $x$-axis relatively far away from each other, without initial velocities. They will attract and then pass through each other before escaping to infinity, as shown in Figure \ref{fig:swapOnce}. The classical simulation with the same initial conditions is plotted for comparison, which is significantly different from the quantum simulation starting from the time of the collision and afterwards. We see that after the collision the charge resides almost exclusively in the mean field. This means that the quantum modes are un-excited in this case, and the charge exchange between the mean field and the quantum modes within a small Q-ball (see Figure \ref{fig:charge-singleQball_2}) is quenched by the collision. It also appears that the charge exchange with the modes does not resume after the collision. This may be related to the resulting charge lumps having charges below the limit of true Q-balls.

\subsection{Charge-swapping Q-balls}
\label{sec:CSQ}

It has been previously pointed out in the classical theory that when Q-balls with opposite charges are placed close to each other, a tower of metastable nonlinear structures or composite Q-balls, dubbed charge-swapping Q-balls (CSQs), can be formed \cite{Copeland:2014qra,Xie:2021glp}. Within a CSQ, positive and negative charges in opposite spatial regions swap with each other periodically, and in 2+1D these structures are very stable \cite{Hou:2022jcd}. In this section, we investigate how the CSQs behave when including quantum corrections. For simplicity, we will focus on the dipole CSQ where there is only one positive charge and one negative charge in the opposite direction. 
\begin{figure}[tbp]
\centering
\includegraphics[width=.45\textwidth]{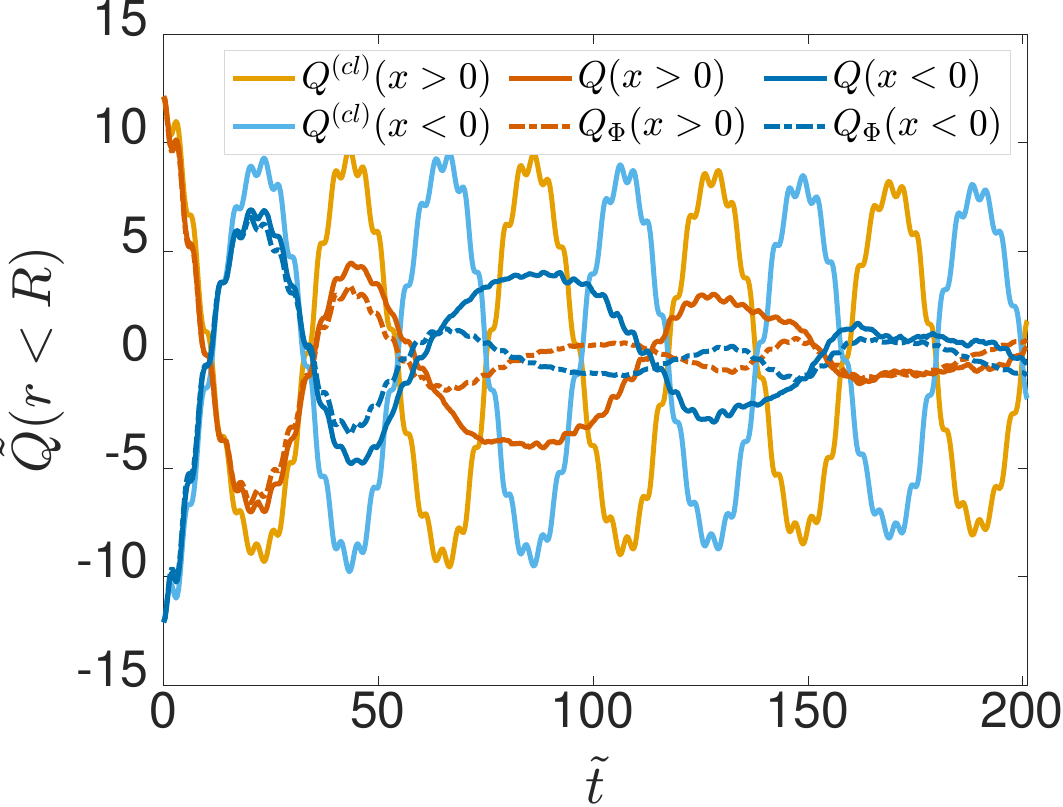}
\hfill
\includegraphics[width=.45\textwidth]{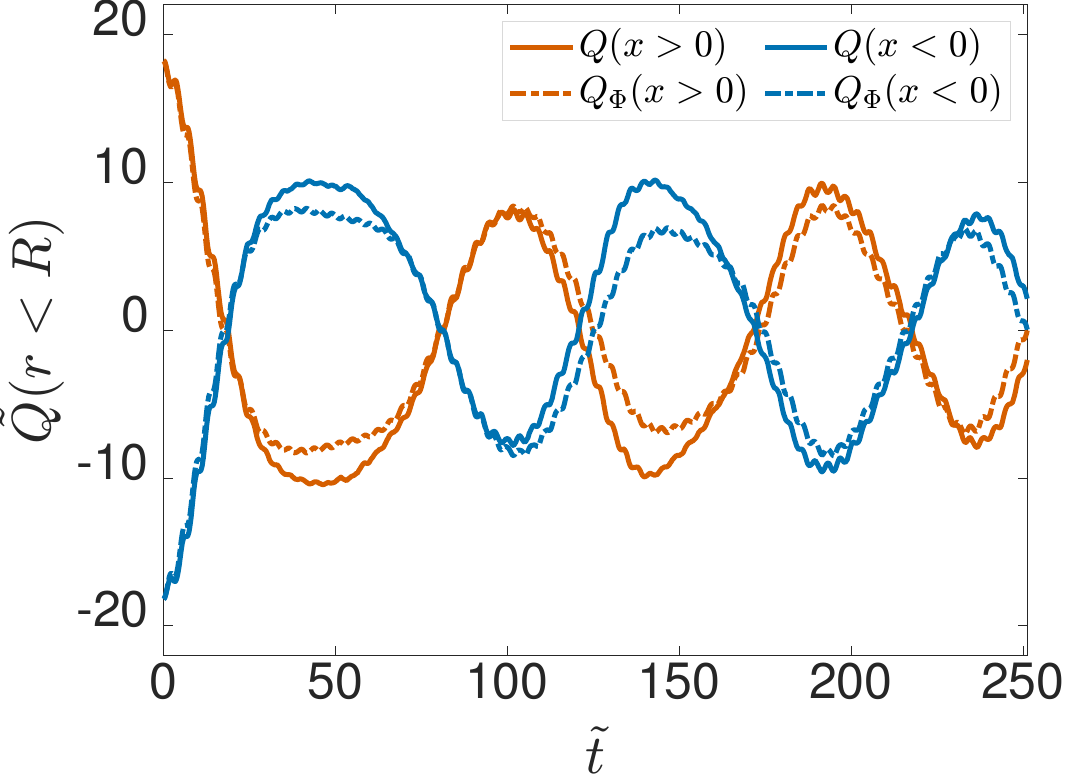}
\caption{\label{fig:charge-csq_wQ0p86}
The evolution of various charges for a dipole Charge-Swapping Q-ball (CSQ) with $|\t{\oi}|=0.86$ and the initial coordinates of two Q-balls being $(\pm1.4,0)/m_r$ (left) and $(\pm3,0)/m_r$ (right). ${}^{(cl)}$ denotes the corresponding classical simulation. The number of sites per dimension is $N=180$, the number of the ensemble realizations is $\mc{E}=20,000$ and $d\t{t}/d\t{x}=0.05$. We only show the charge within a radius of $\t{R}=17$ from the origin. 
}
\end{figure}

A dipole CSQ can be prepared by relaxing a system of two superimposed Q-balls placed tightly together such that their nonlinear cores overlap. In Figure \ref{fig:charge-csq_wQ0p86}, we compare the evolution of a dipole CSQ in the classical and quantum simulation, from the point view of the charge within a radius of $\t{R}=17$ from the origin. The initial Q-balls are placed on the $x$-axis symmetrically about the origin. To calculate the charge of of the CSQ, we integrate the charges in a half disk centered at the origin for both $x>0$ and $x<0$. The classical simulation reveals a consistent charge-swapping pattern in the steady state, and the system experiences a gradual dissipation of energy. When the quantum effects are included, the evolution pattern is quite different. If the Q-balls are placed in the same positions as the classical case, the composite structure dissipates much faster than the classical case, and we see that the mean field fails to be a good approximation of the total field after the first swap of charges. That is, the quantum mode two-point functions pick up a significant fraction of the charge after the two Q-balls strongly interact several times. 

In the quantum case, to produce a more stable CSQ, we need to increase the initial distance between the two constituent Q-balls, as shown for the case of an initial separation of $(\pm3,0)/m_r$ in Figure \ref{fig:charge-csq_wQ0p86}. In this case, the two oscillating lumps have greater velocities when their cores overlap and interact strongly. This is somewhat similar to the collision case in the last subsection where the quantum corrections can be effectively quenched. However, in this case, the swapping period of the CSQ is greater than the classical counterpart. As it is computationally much more expensive to perform the quantum simulations, we are only able to evolve the quantum CSQ to a fraction of time of the classical simulations. To simulate the system more reliably for a longer time, we need to avoid the unphysical waves from the periodic boundaries. To this end, we use a larger simulation box with a length of $251.2/m_r$, keeping the same lattice spacing. For this box, the light crossing time, which is the least time for a site to have a causal connection with itself, is equal to the box length, during which time the system in the center will not be affected by the unphysical propagating waves. For these slightly longer simulations, the results turn out to be similar.

\begin{figure}[tbp]
\centering
\includegraphics[width=.5\textwidth]{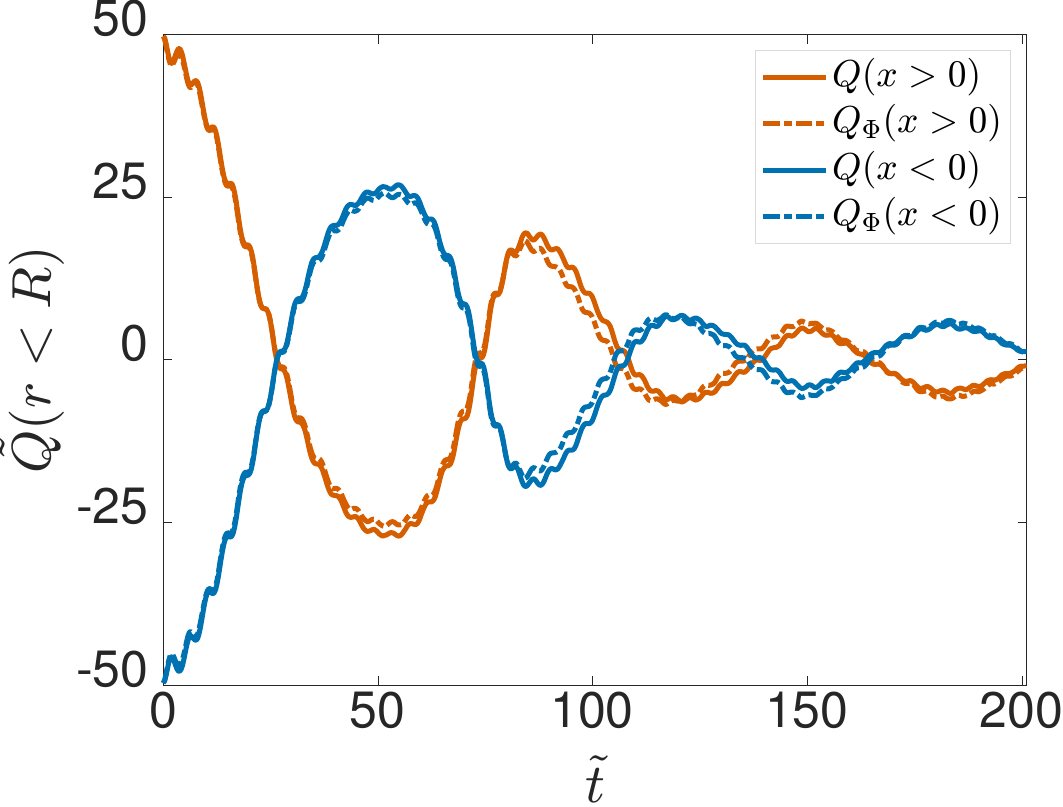}
\caption{\label{fig:charge-csq_wQ0p78}
Quantum corrected charge-swapping Q-balls with two large constituent Q-balls ($\t{\oi}=\pm0.78$), initially placed at $(\pm3,0)/m_r$. The other numerical setup is the same as in Figure \ref{fig:charge-csq_wQ0p86}. 
}
\end{figure}

The situation is again similar when the initial constituent Q-balls are of the thin-wall type, as can be seen in Figure \ref{fig:charge-csq_wQ0p78}. In this case, although the mean fields provide the dominant contributions, one still needs to initially have a larger separation for the constituent Q-balls for a more stable CSQ to form, and the initial oscillating lump typically sheds away more energy in the initial relaxation process. 

\begin{figure}[tbp]
\centering
\includegraphics[width=.46\textwidth]{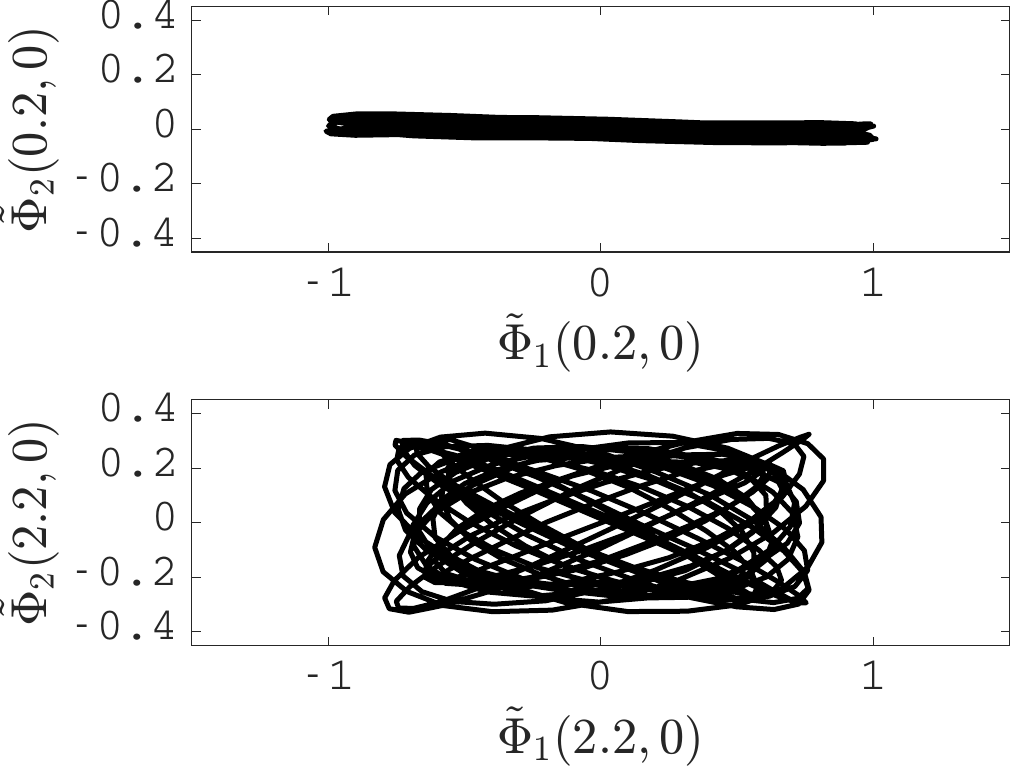}
\hfill 
\includegraphics[width=.45\textwidth]{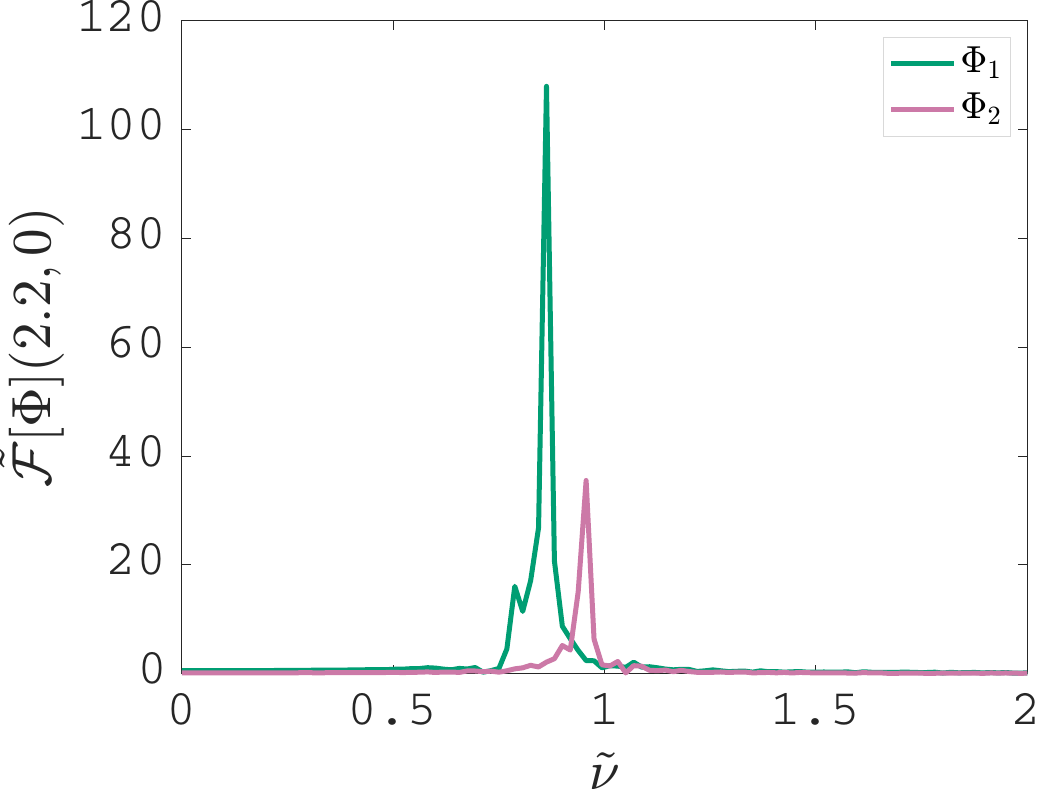}
\caption{\label{fig:phasePlane-csq}
Phase portraits in the $\Phi_1$-$\Phi_2$ plane for two representative points for the time duration $t=(166\sim355)/m_r$. The right plot shows the Fourier transforms of the one-point fields $\Phi_j(t)$ at point $(2.2,0)/m_r$ during $t=(166\sim500)/m_r$. The initial two opposite Q-balls with $\t{\oi}=\pm0.86$ has a distance of $5/m_r$. The number of sites per dimension is $N=628$, the number of fields in the ensemble realization is $\mc{E}=5,000$ and $d\t{t}/d\t{x}=0.1$. 
}
\end{figure}
Let us consider the evolution of a couple of representative points of the CSQ in the quantum simulations. In Figure \ref{fig:phasePlane-csq}, we plot the phase portraits of the mean field at two spatial points in the $\Phi_1$-$\Phi_2$ plane, using the data of about $5$ charge-swapping periods after the CSQ is properly formed. The results are very similar to the classical simulations: the orbit is an approximate ellipse that rotates ever so slightly in each oscillation, filling a rectangle area that becomes more square-like as the point moves away from the center of the CSQ. As in the classical case, the charge-swapping period can be very precisely estimated from the difference between the dominant frequencies \cite{Xie:2021glp}. For Figure \ref{fig:phasePlane-csq}, the dominant frequencies are respectively $\t{\oi}_1=0.86$ and $\t{\oi}_2=0.96$, which gives a swapping period of $67/m_r$.

\begin{figure}[tbp]
\centering
\includegraphics[width=.5\textwidth]{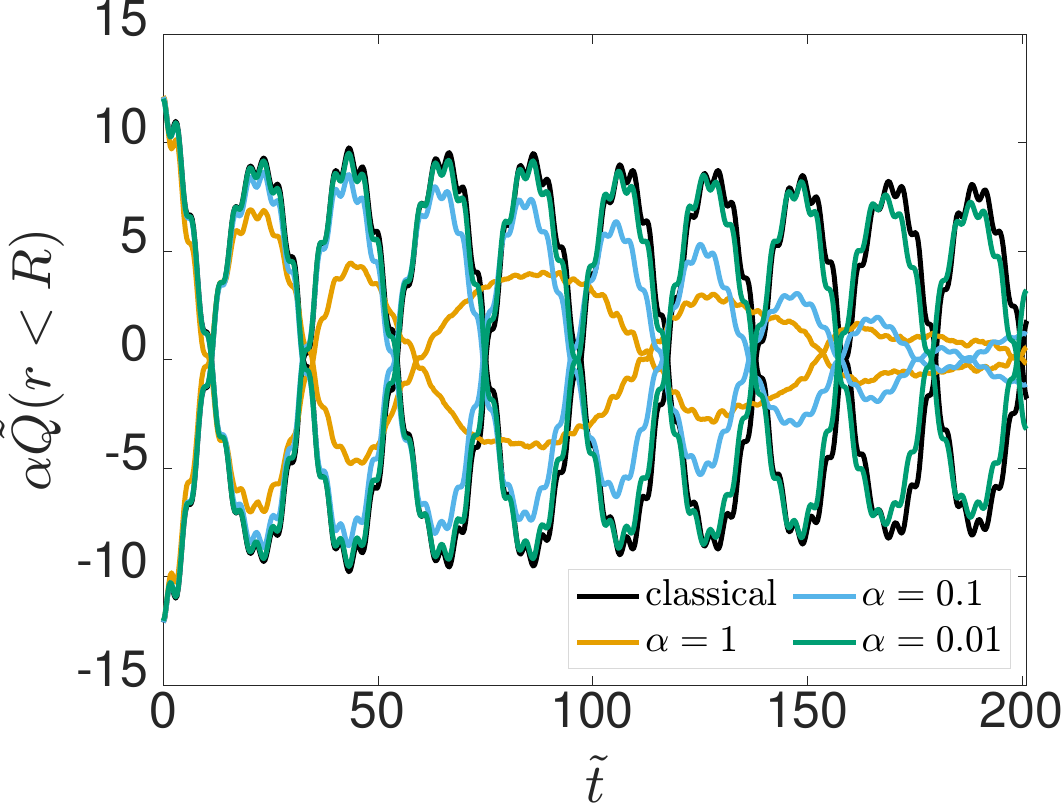}
\caption{\label{fig:csq-diffLambda}
Comparison of CSQ charge evolution from simulations with different rescalings $\alpha$. The two Q-balls, with $\t{\oi}=\pm0.86$, are placed at $(\pm1.4,0)/m_r$ initially. We only show the charge within a radius of $\t{R}=17$ from the origin. The opposite lines with the same color are for the region of $x<0$ and $x>0$ respectively.
The number of sites per dimension is $N=180$ and $d\t{t}/d\t{x}=0.1$. The number of fields in the ensemble realization is $\mc{E}=20,000$ for quantum cases.
}
\end{figure}

Finally, the results in this subsection involve the fiducial values for the potential, where the Q-ball is fairly small or conversely the quantum corrections are relatively large. Again applying the rescaling procedure leading to Figure \ref{fig:charge-singleQball_2}, we can recover the classical results for the CSQs. In Figure \ref{fig:csq-diffLambda}, we show how the CSQs including quantum corrections reduces to the corresponding classical CSQs, as the rescaling parameter $\alpha$ is decreased. Similar conclusions apply for well-separated Q-balls and collisions of Q-balls.

\section{Summary and conclusions}
\label{sec:summary}

In this work, we investigate the dynamics of Q-balls including quantum corrections in a $U(1)$ scalar field theory with a $\cph^6$ potential. The quantum evolution is implemented using the inhomogeneous Hartree approximation, improved by the statistical ensemble method to enhance simulation speed. 

We find that there are two evolution patterns for a single Q-ball in 2+1D. For large Q-balls (large charge), the mean field evolution closely resembles the purely classical simulations and the quantum modes remain largely un-excited. This aligns with the consensus that classical dynamics is a good approximations to the fully quantum dynamics when the occupation numbers of the relevant momentum modes are high. For small Q-balls (small charge), the quantum modes are excited and swap a large fraction of the total charge periodically with the mean fields. This is also expected as when the occupation numbers are low the quantum effects should be significant. However, the exact manifestation can only be attained via numerical simulation. In contrast to Q-balls in 3+1D, which for the fiducial $\mathcal{O}(1)$ parameters are unstable under quantum corrections for a big portion of the classically stable frequency range, we find that Q-balls in 2+1D remain stable at all frequencies. Through a simple rescaling, we can find larger-amplitude Q-balls, for which the quantum corrections are arbitrarily small.

We also make first attempts at exploring the quantum corrections to the dynamics between multiple Q-balls and study three scenarios, focusing on the differences from purely classical dynamics. For two identical, well-separated Q-balls with a phase difference being $\pi/2$, the quantum effects do not modify the charge transfer rate between the two Q-balls but have some influence on the end states. Small Q-balls exhibit slightly larger velocities, which are produced by the forces between the Q-balls, while large Q-balls basically keep the same velocities. When two well-separated Q-balls with opposite charges and no initial charge in the quantum modes undergo a fast collision and escape, we find the charge exchange between the mean fields and the perturbations may be quenched. We have also considered dipole CSQs, bound states formed from two Q-balls with opposite charges. These highly non-linear and composite structures persist under quantum corrections, in which case the main features of CSQs are retained, but they decay faster and the two original Q-balls being further apart initially can make the CSQ more stable after the relaxation process. On the other hand, scaling up the CSQs again makes the quantum corrections negligible.

Our findings indicate that, within the inhomogeneous Hartree approximation, the dynamics of a system featuring multiple Q-balls is qualitatively similar to the classical system. While the Hartree approximation is in a sense only the lowest order truncation of the Dyson-Schwinger hierarchy, for inhomogeneous systems it can provide information on quantum decay channels and transient thermalisation \cite{Salle:2000hd,Salle:2003ju}. Further exploring not only where the charge, but also the energy, goes in Q-ball collisions and decay processes would be of great interest. Another intriguing (and highly ambitious) avenue for future exploration involves going to NLO order in a 2PI expansion \cite{Berges:2000ur,Aarts:2001qa,Berges:2004yj,Arrizabalaga:2005tf}. For inhomogeneous systems, this has to our knowledge not been done in 2+1D or 3+1D, and represents a further level of numerical complexity as it involves memory integrals over the past history of the system.

\acknowledgments

SYZ acknowledges support from the Fundamental Research Funds for the Central Universities under grant No.~WK2030000036, from the National Natural Science Foundation of China under grant No.~12075233 and 12247103, and from the National Key R\&D Program of China under grant No. 2022YFC220010. The work of PMS was supported by an STFC Consolidated Grant [Grant No.~ST/T000732/1],


\bibliographystyle{JHEP}
\bibliography{refs}






\end{document}